\documentclass[]{spie}  

\usepackage{amsmath,amsfonts,amssymb}
\usepackage{graphicx}
\usepackage[colorlinks=true, allcolors=blue]{hyperref}
\usepackage{threeparttable}
\usepackage{array} 

\title{Automatic detection and characterization of random telegraph noise in sCMOS sensors}

\author[a]{Arda Özdoğru}
\author[b]{Sergey Karpov}
\author[b]{Asen Christov}
\author[a]{Stanislav Vítek}
\affil[a]{Czech Technical University in Prague, Jugoslávských partyzánů 1580/3, 160 00, Prague, Czechia}
\affil[b]{Institute of Physics of the Czech Academy of Sciences, Na Slovance 1999/2, 182 00, Prague, Czechia}

\authorinfo{Further author information: (Send correspondence to A.Ö.)\\A.Ö.: E-mail: ozdogard@cvut.cz \\  S.K.: E-mail: karpov@fzu.cz, Telephone: +420 
266 05 2135 \\ A.C.: E-mail: christov@fzu.cz, Telephone: +420 
266 05 2135 \\ S.V.: E-mail: viteks@fel.cvut.cz, Telephone: +420 224 35 2232 }

\begin{document} 
\maketitle

\begin{abstract}
Scientific CMOS (sCMOS) image sensors are a modern alternative to typical CCD detectors and are rapidly gaining popularity in observational astronomy due to their large sizes, low read-out noise, high frame rates, and cheap manufacturing. However, numerous challenges remain in using them due to fundamental differences between CCD and CMOS architectures, especially concerning the pixel-dependent and non-Gaussian nature of their read-out noise. 
One of the main components of the latter is the random telegraph noise (RTN) caused by the charge traps introduced by the defects close to the oxide-silicon interface in sCMOS image sensors which manifests itself as discrete jumps in a pixel’s output signal, degrading the overall image fidelity. In this work, we present a statistical method to detect and characterize RTN-affected pixels using a series of dark frames. Identifying RTN contaminated pixels enables post-processing strategies that mitigate their impact and the development of manufacturing quality metrics.
\end{abstract}


\section{INTRODUCTION}
\label{sec:intro}  

Scientific complementary metal–oxide–semiconductor (sCMOS) detectors have emerged as a compelling alternative to traditional charge-coupled device (CCD) technology in high-performance astronomical imaging. Their attributes of large sensor formats, low read noise~\cite{Fowler10}, high frame rates, and low fabrication costs offer substantial advantages for sky survey applications~\cite{Karpov20}. Yet, the sCMOS architecture introduces challenges that differ fundamentally from CCDs. One of the most prominent among these is the pixel-dependent, non-Gaussian read-out noise. 

A principal contributor to this non-Gaussian behavior is random telegraph noise (RTN), arising from defects near the oxide–silicon interface forming charge traps~\cite{Campbell08}. These traps cause discrete “telegraph”-like jumps in pixel outputs due to the capture and emission of electrons. These traps are mostly in the oxide layer and distributed into different depths exerting different probabilities, hence forming a complex structure of possible jumping/tunneling mechanisms, each contributing to RTN\cite{Vecchi23}. This type of noise presents itself in all semiconductor materials~\cite{Simoen16}. For array-type semiconductor devices such as random access memories and image sensors, even though the percentage of cells that are affected by RTN is not high~\cite{Chao2017}, its effect on decreased yields and image fidelity, respectively, cannot be denied. Especially in the sCMOS detectors where the read-out noise (RON) is as low as a couple of electrons, RTN biases the measurement of faint sources and harms the reliability of observed data. Therefore, characterizing RTN is of high interest in areas such as hardware development, fault correction, sky-survey applications, and image data post-processing methods.

There are many studies focusing on the circuit-level characterization of RTN with methods such as a variety of statistical analysis~\cite{Wang06, Chao2017, Joe11, Islam17}, and Hidden Markov Model-based probabilistic characterizations~\cite{Ito11, Puglisi2013}. However, circuit-level characterizations, with or without the help of further external complex electronic systems, cannot be reproduced nor applied by the end consumers and, therefore, have a limited reach compared to their importance. Moreover, analysis that relies on RMS of pixel noise measurements fell short especially when during sensor-wide characterization because in histograms RTN contaminated pixels present themselves in an interleaving manner with the so-called hot pixels which by themselves exhibit higher standard deviations. Hence, creating a system-level characterization of RTN for each pixel individually based on the final images produced by either the image sensor raw data, or the camera system output carries a further reaching importance that can be applied, not only theoretically but also practically aiding efforts to harness sCMOS’s high-speed and large-format benefits for precision astronomical measurement.

\begin{figure} [ht]
\begin{center}
\begin{tabular}{c}
\includegraphics[height=10cm]{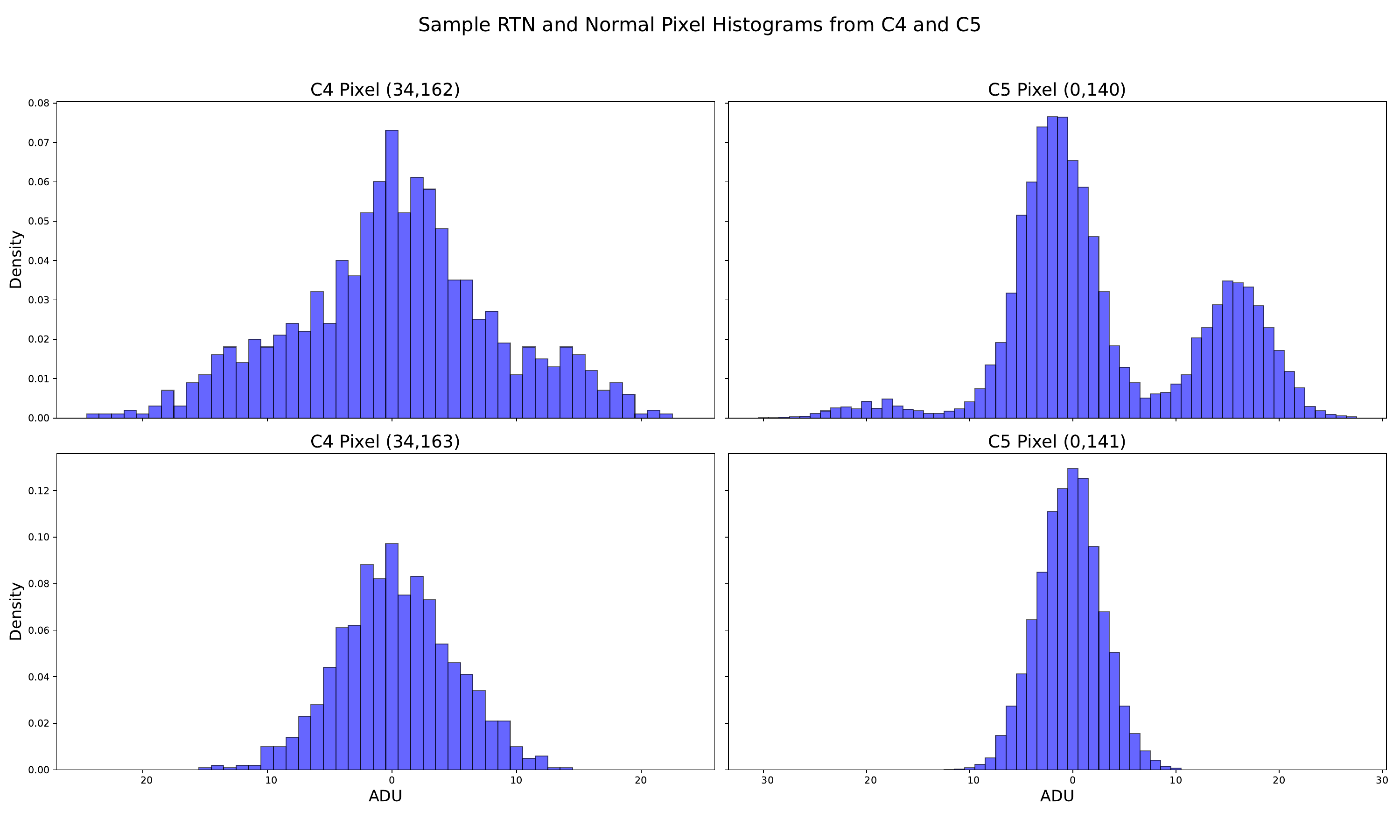}
\end{tabular}
\end{center}
\caption
{ \label{fig:hist_c4_c5} Example pixel histograms. Top row: RTN contaminated pixels. Bottom row: normal pixels. Left column is for camera C4. Right column is for camera C5.}
\end{figure}

In this work, we develop a statistical algorithm for detecting and characterizing RTN-afflicted pixels from a series of dark frames. The approach is based on the assumption that any pixel that RTN does not contaminate acquires a Gaussian distribution over a sufficient amount of data points and whenever RTN is present, the pixels deviate from Gaussian distribution. In the most definitive cases of RTN contamination, the time series histogram of these pixels approach multiple, in most cases three or more~\cite{Chen14}, distinct approximately Gaussian distribution, so a mixture of Gaussians. Example pixels are shown in Fig.~\ref{fig:hist_c4_c5}.

The paper is structured in the following way. In Sec.~\ref{sec:sensor}, the camera and sensor properties related to our experiments are introduced and the experiment settings are noted. Sec.~\ref{sec:algo} defines the algorithm that is developed for RTN detection. The detection results and further statistics about the RTN in the sensors are presented in Sec.~\ref{sec:res}. A conclusion and the future work are laid out in Sec.~\ref{sec:conc}.

\section{Experiment Parameters}\label{sec:sensor}

Images obtained from the two different cameras incorporating sCMOS sensors, namely Moravian Instruments C4-16000EC and Moravian Instruments C5A-100M, are analyzed. From here on out, these cameras are mentioned as C4 and C5, respectively. The properties of both cameras are presented in Tab. \ref{tab:sensors}.

\begin{table}[t]
\centering
\begin{threeparttable}
\caption{Comparison of C4-16000 (EC) and C5A-100M sCMOS Sensors}
\label{tab:sensors}
\begin{tabular}{|p{3.5cm}|p{6cm}|p{6cm}|}
\hline
\textbf{Parameter} & \textbf{C4-16000 (EC)} & \textbf{C5A-100M} \\
\hline
Sensor Model & Gpixel GSENSE4040  & Sony IMX461 \\
\hline
Effective Resolution & 4096 $\times$ 4096 (16 MP)  & 11,664 $\times$ 8,750 ($\sim$102 MP) \\
\hline
Pixel Pitch & 9 $\mu$m $\times$ 9 $\mu$m    & 3.76 $\mu$m $\times$ 3.76 $\mu$m \\
\hline
Active Area & $\sim$36.9 $\times$ 36.9 mm    & $\sim$43.9 $\times$ 32.9 mm \\
\hline
Dark Pixels & Right 64 px & Left 44 px, Right 44 px\\
           &  & Top 38 px, Bottom 38 px \\
\hline
Gain Architecture & Dual 12-bit channels (high/low gain) combined 16-bit HDR output  & Full 16-bit linear digitization \\
\hline
Read-out Noise & 3.7$e^-$ & 2.7$e^-$ \\
\hline
Cooling (below ambient) & $\sim$30–35$^\circ$C  & $\sim$40-45$^\circ$C \\
\hline
Noise Suppression & Correlated Double Sampling & Advanced Error Correction\tnote{1} \\
\hline
Output ADU datatype &  16-bit Int  &  16-bit Int \\
\hline
\end{tabular}
\begin{tablenotes}
\small
\item[1] No detailed information was found about the underlying mechanism.
\end{tablenotes}
\end{threeparttable}
\end{table}

\begin{table}[h]
\centering
\caption{Image acquisition parameters}
\label{tab:image_param}
\begin{tabular}{|l|c|c|}
\hline
\rule[-1ex]{0pt}{3.5ex}  \textbf{Parameter} & \textbf{C4} & \textbf{C5}  \\
\hline
\rule[-1ex]{0pt}{3.5ex}  Number of Images   & 1000       & 10000       \\
\hline
\rule[-1ex]{0pt}{3.5ex}  Sensor Temperature ($^\circ\text{C}$) & $-10$ & $-20$ \\
\hline
\rule[-1ex]{0pt}{3.5ex}  Exposure Time (s)     & 0.01     & 5         \\
\hline
\end{tabular}
\end{table}

The dark images are obtained with closed shutters in a dark box setup prepared for further examination of the cameras in a dark lab. Image acquisition is done differently for each of the cameras. The settings for image acquisitions are shown in Tab. \ref{tab:image_param}.
The reason for the temperature difference is caused by the inherent cooling in the cameras and the lab temperature which was between $15$ - $20\,^\circ\text{C}$. During the initial tries, it was seen that C4 can hold a stable temperature around $-10\,^\circ\text{C}$ during the capture and read-out processes over long sequences, while C5 was capable of holding a stable $-20\,^\circ\text{C}$.
On the other hand, the reason for exposure time was dependent on the image size and the insufficient read-out computer speed. Due to the image size of C5 ($\sim$102MP), it was seen that for the shutter speeds below 5 seconds, the computer could not read and save the images stably over long sequences. Such a problem was not observed for C4. 
The difference in the number of images was due to the limited availability of the C4 sensor. 

\section{Algorithm}\label{sec:algo}

\begin{figure}[htbp]
    \centering
    \begin{center}
    \begin{tabular}{c} 
    \includegraphics[width=0.7\textwidth]{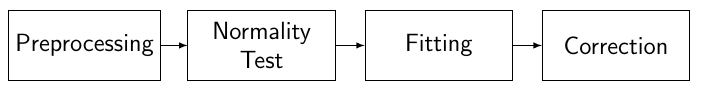}
    \end{tabular}
    \end{center}    
    \caption{\label{fig:block} Block diagram of the RTN detection process.}
\end{figure}

The detection algorithm consists of several steps shown in Fig.~\ref{fig:block}. These steps are thoroughly explained in their related subsections. All the related codes are published in an online repository~\cite{github}.

\subsection{Preprocessing}\label{sec:outlier}

The images, by cameras' defaults, are saved by unsigned 16-bit integers.  After the sequences of dark frames are read, data is converted to 64-bit float for further processing. A form of sigma-clipping is applied to the pixels with a single iteration to reject the extreme values caused by cosmic ray hits, as shown in Eq.~\ref{eq:clip}.

\begin{equation}\label{eq:clip}
px_{\text{new}} =
\begin{cases}
med(px_{frames}), & \text{if } \left| px - med(px_{frames}) \right| > 10\,\sigma_{px_{frames}}, \\
px, & \text{otherwise.}
\end{cases}
\end{equation}
where $px$ denotes the pixel's value in a frame, $med(px_{frames})$ denotes the pixel's median over all the frames, and $\sigma_{px_{frames}}$ is that pixel's standard deviation across all the frames. 

After that, the pixels are centered around 0 through median subtraction. 

\subsection{Normality Test}\label{sec:normality}
Considering the nature of our data and our preliminary examinations, we have observed that an "ideal" pixel that is minimally contaminated with noise would result in output values that depend on the thermal generation of electrons for a given temperature and the exposure time, which results in a bias value and a charge that accumulates due to dark current over the exposure time. The dark current is negligibly low compared to the bias values in the experimented sensor temperatures and the exposure times. However, due to the RON present in the readout electronics of sCMOS sensors, the output of a pixel spreads into a normal distribution. 

\begin{figure} [ht]
\begin{center}
\begin{tabular}{c} 
\includegraphics[height=8cm]{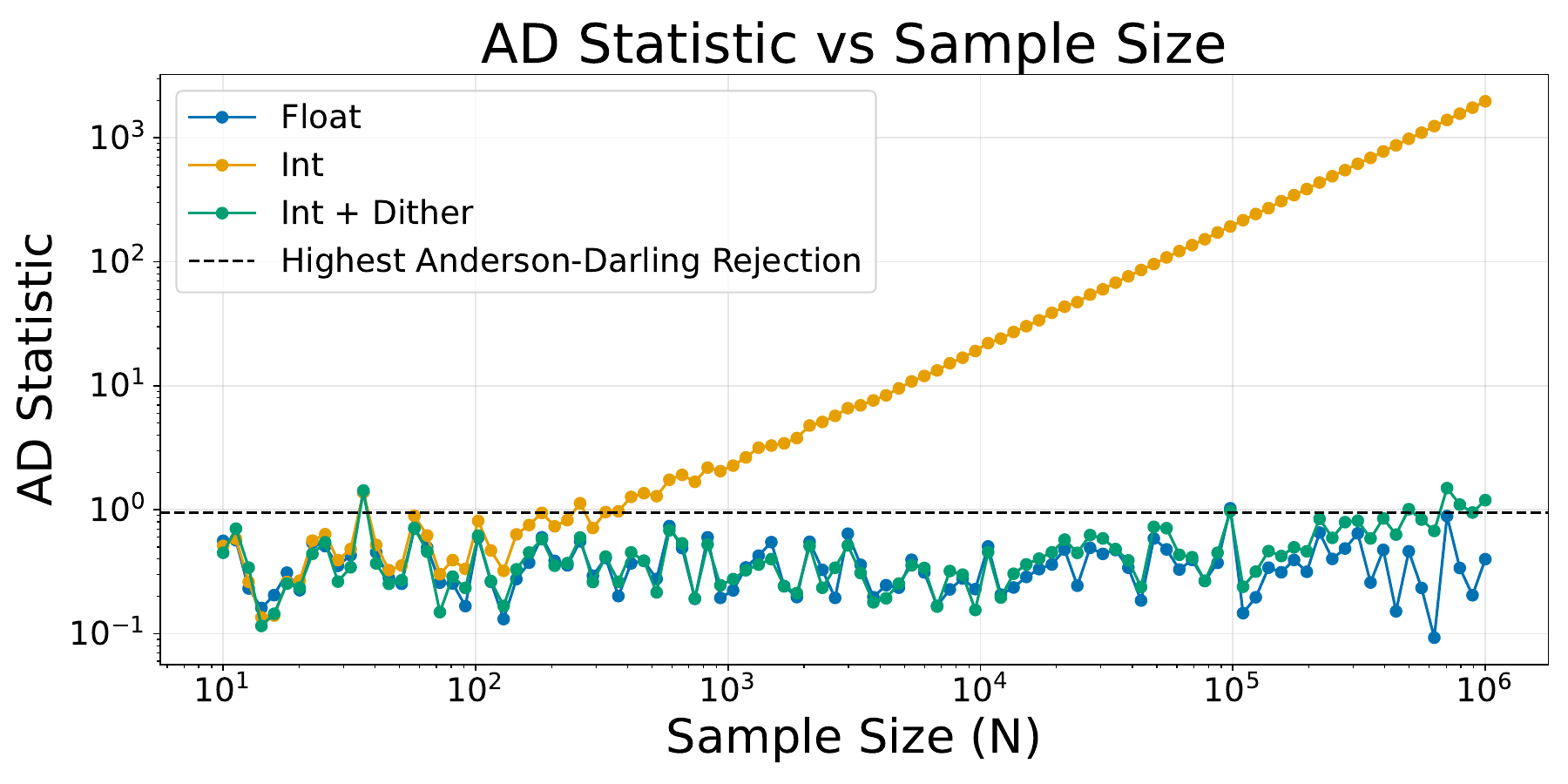}
\end{tabular}
\end{center}
\caption
{ \label{fig:ad_dither} AD statistics for float, rounded integer, and rounded integer with dithered sampling from the artificial normal distributions over a range of sample sizes.}
\end{figure} 

In state-of-the-art sCMOS sensors, RTN appears in around 1-2\% of the pixels~\cite{Chao2017,Leyris06}. Given the amount of pixels and the data points, directly fitting each pixel becomes computationally expensive. In order to reduce total computational costs and create an initial classification, the Anderson-Darling normality test, sensitive to deviations in the tails, is adopted.
Normality tests, in general, are practical and accurate tests for detecting deviations from normal distribution for a sufficient amount of data which is generally around hundred data points~\cite{Razali10}. 

As previously mentioned, a pixel contaminated by RTN is expected to deviate from normal distribution and approach a distinctly characterizable mixture of Gaussians due to the discreet jumps caused by RTN and the spread caused around each level due to the RON. Therefore, any pixel that contains a detectable RTN is expected to be rejected by the Anderson-Darling (AD) normality test. 

In order to check the stability and viability of the AD statistics, artificial data with normal distribution is created for different sample sizes. It was discovered that AD statistics behave differently for integer rounded data, which represents our dataset, and the float sampling of the artificial normal distributions. This is caused by the repeated values that occur in the dataset due to rounding, which classifies the distribution as non-normal even though it is. To overcome this issue, a slightly modified version of dithering, widely used in the field of audio compression, is adopted. This method is shown to effectively reduce quantization noise that occurs during audio compression~\cite{Schobben95}. The method is adopted in the following manner: each data point is summed up by a number randomly selected from a uniform distribution between $[-0.5, 0.5)$ before calculating the AD statistic. 
The results of AD statistics for float, rounded integer, and rounded integer with dithered sampling from the artificial normal distributions for different sample sizes are shown in Fig.~\ref{fig:ad_dither}. It can be observed that with growing sample sizes, while the integer rounding creates an exponential rise of AD statistics rejecting the normal distributions, the dithered version results in scores that are comparable with the float-based sampling, until a very high number of data points (\(\sim \)$10^6$). Considering the scope of the paper, where between $10^3 - 10^4$ data points are used, this method serves as a viable and practical approach.

\textit{An important mention is that the dithering is only used for the AD statistics calculation and is not incorporated into the original data for any other step in our processing pipeline.}

In terms of classification, the pixels that are rejected by all significance levels of the AD test for normality are chosen for further processing. Even though this early classification is made, it does not necessarily mean that all the pixels rejected by the AD test necessarily contain RTN.

\begin{figure} [ht]
\begin{center}
\begin{tabular}{c}
\includegraphics[height=8cm]{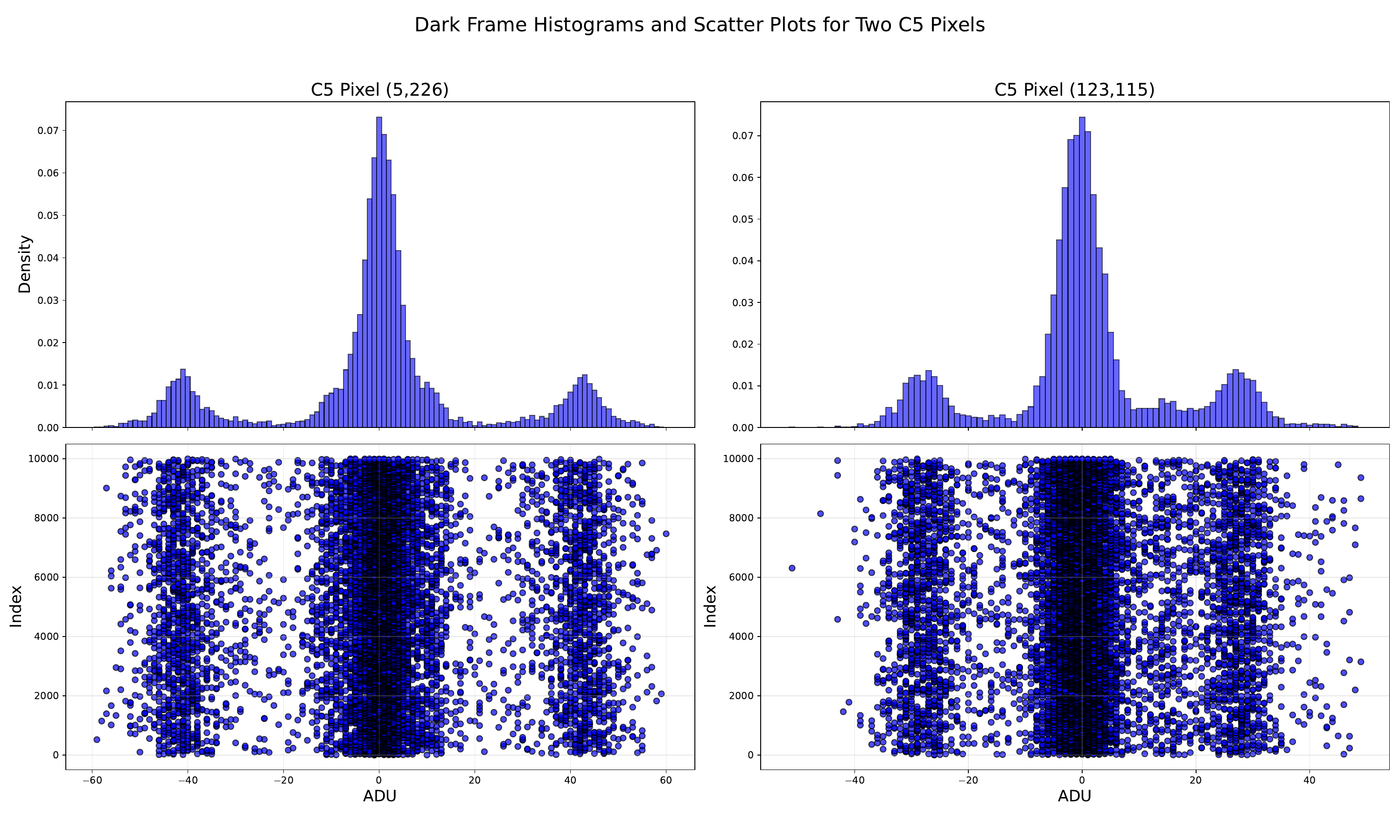}
\end{tabular}
\end{center}
\caption
{ \label{fig:c5_hist_scatter} Two example pixels from C5, potential candidates for mixture of 5 Gaussians.}
\end{figure}

\subsection{Fitting}\label{sec:fitting}

Pixels rejected by the AD normality test are fitted with a method adjusted based on the preliminary visual examination of the obtained data and the assumptions made earlier, where an "ideal" pixel exhibits a normal distribution around its bias value over a range of images and a pixel contaminated by RTN exhibits distinctly identifiable multiple normal distributions. 

A preliminary visual inspection of the pixels of a sensor is needed to determine rough initial and boundary parameters for the fitting. A more accurate estimation of the initial parameters will result in a better fit, therefore improving the detection quality. 

Due to the expected behavior of RTN, the fitting function utilizes a mixture of Gaussians. The maximum amount of Gaussians included in the mixture is determined during the preliminary visual inspection. This is mainly to decrease overfitting the data with an unnecessary higher level mixture. An example of two pixels from C5 with a potential mixture of 5 Gaussians is shown in Fig.~\ref{fig:c5_hist_scatter}.

The fitting method is as follows. The histograms of the pixel output over the sequence of images, where the bins are adjusted to encapsulate individual integer values between and including the minimum and the maximum of the pixel values, are created. The resulting bin densities with the corresponding bin centers are fitted to single Gaussian (1G), three (3G) and five Gaussian (5G) mixtures using $curve\_fit$ function from SciPy library~\cite{curve_fit}. 

\begin{figure}[b]
  \centering
  \begin{tabular}{ccc}
    \includegraphics[width=0.3\textwidth]{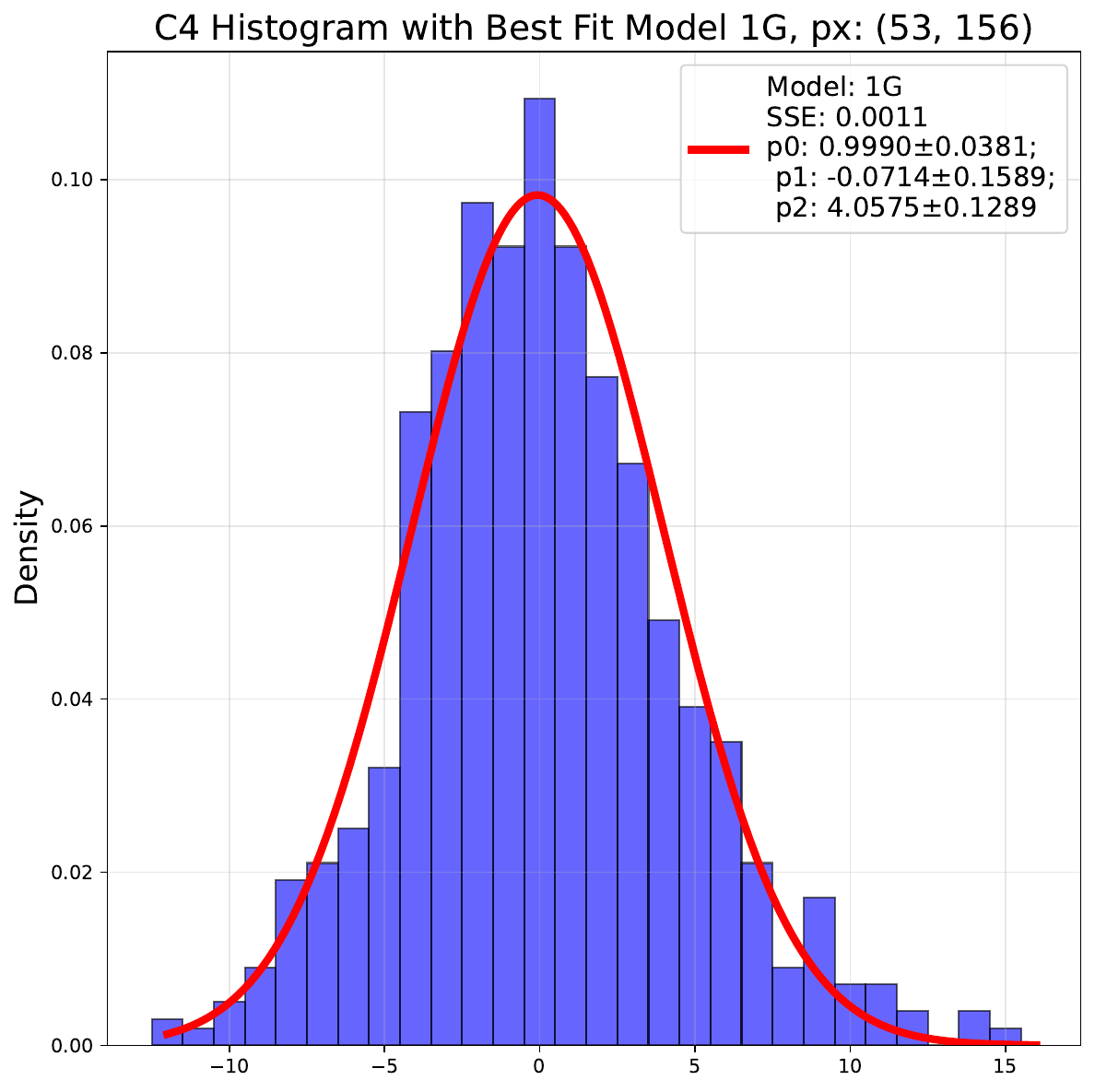} & 
    \includegraphics[width=0.3\textwidth]{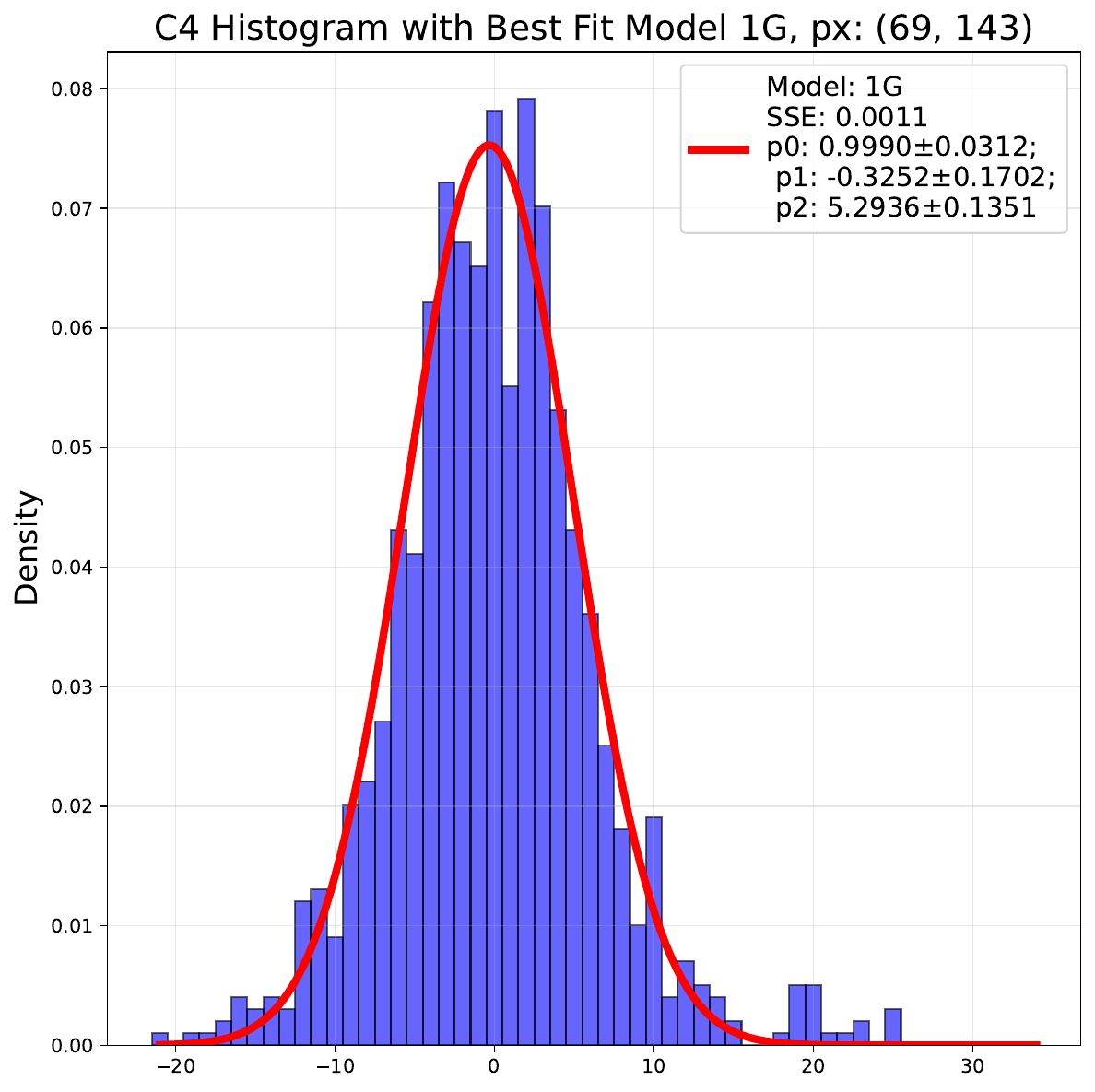} & 
    \includegraphics[width=0.3\textwidth]{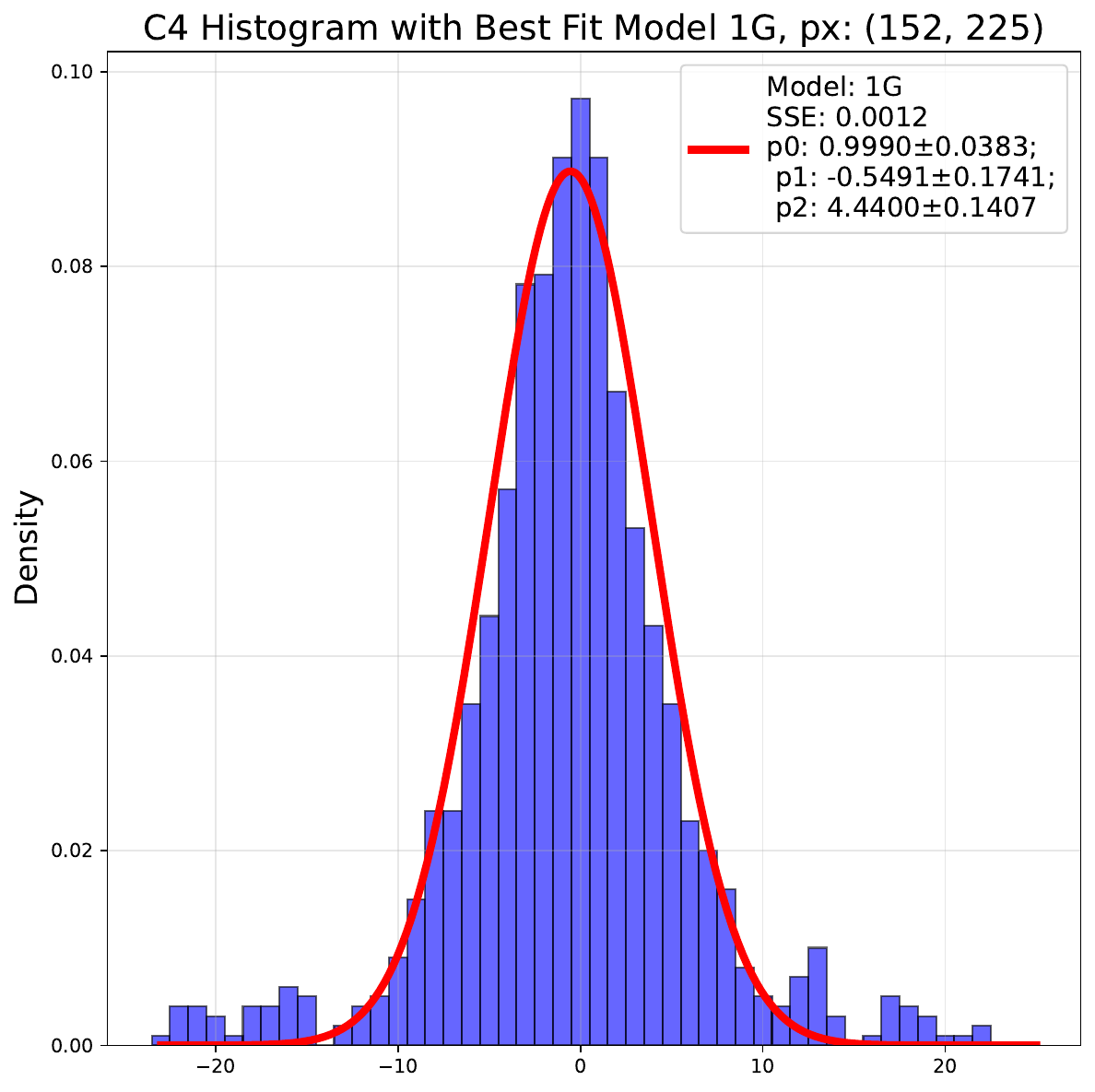} \\
    \includegraphics[width=0.3\textwidth]{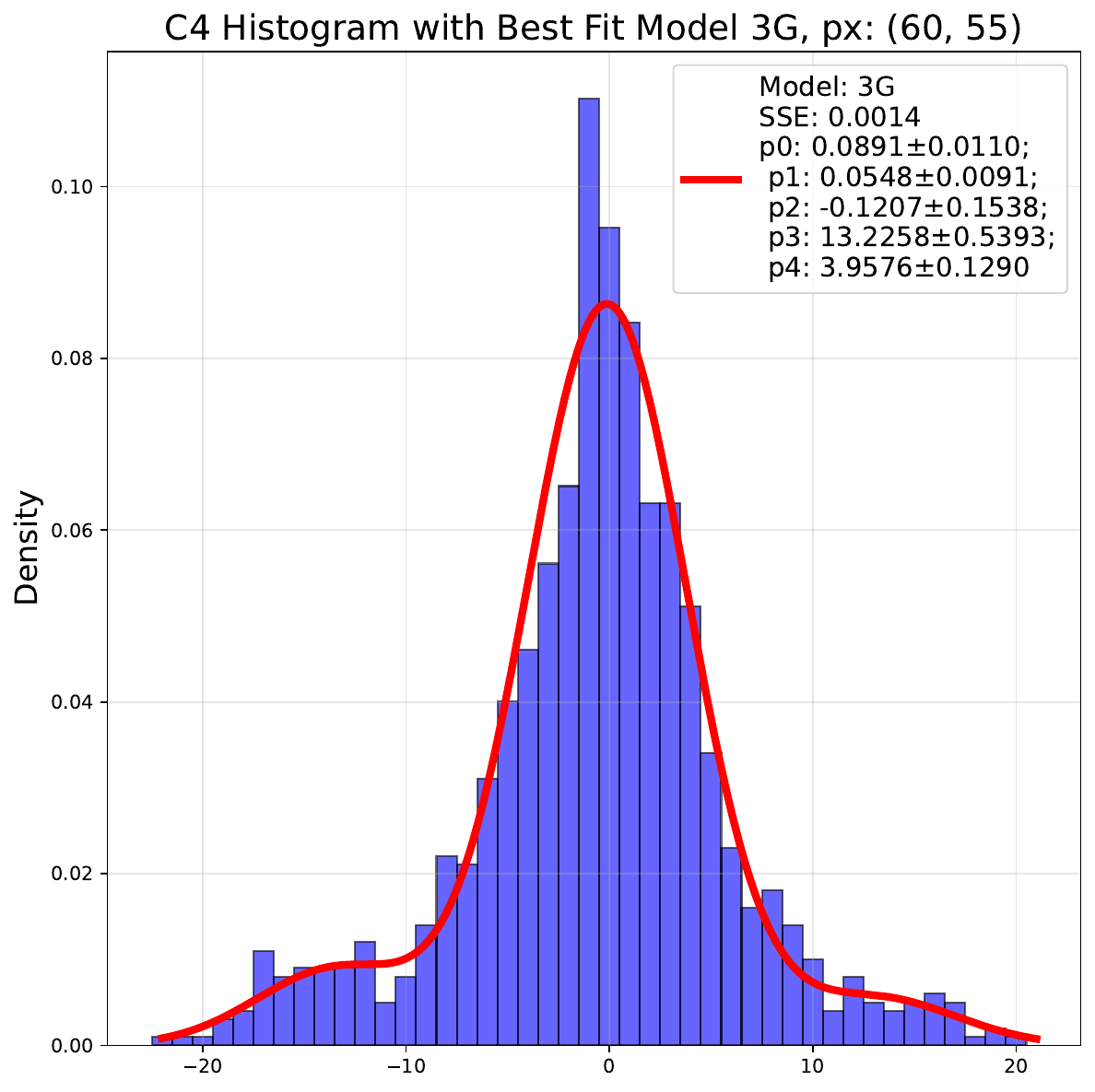} & 
    \includegraphics[width=0.3\textwidth]{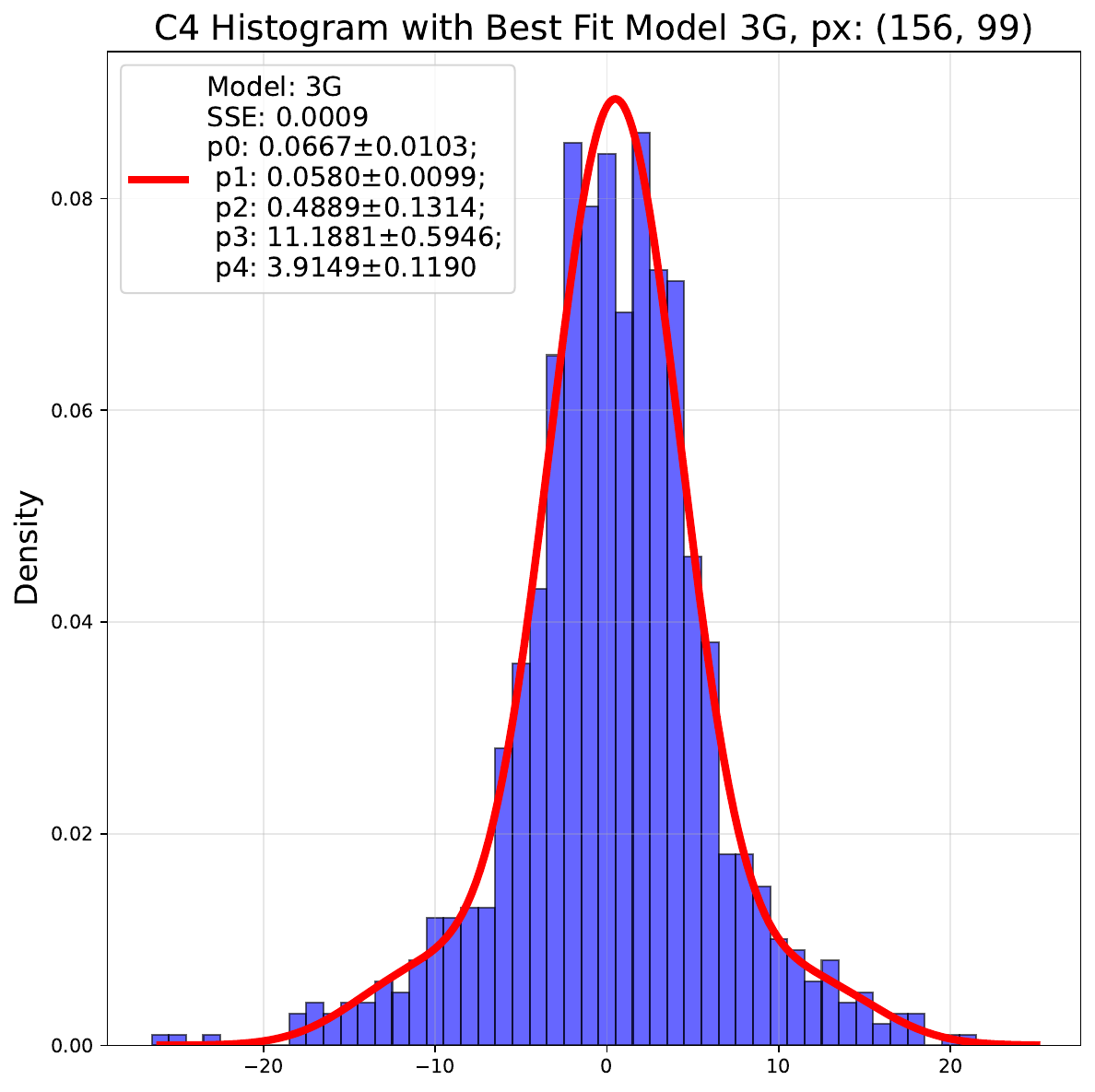} & 
    \includegraphics[width=0.3\textwidth]{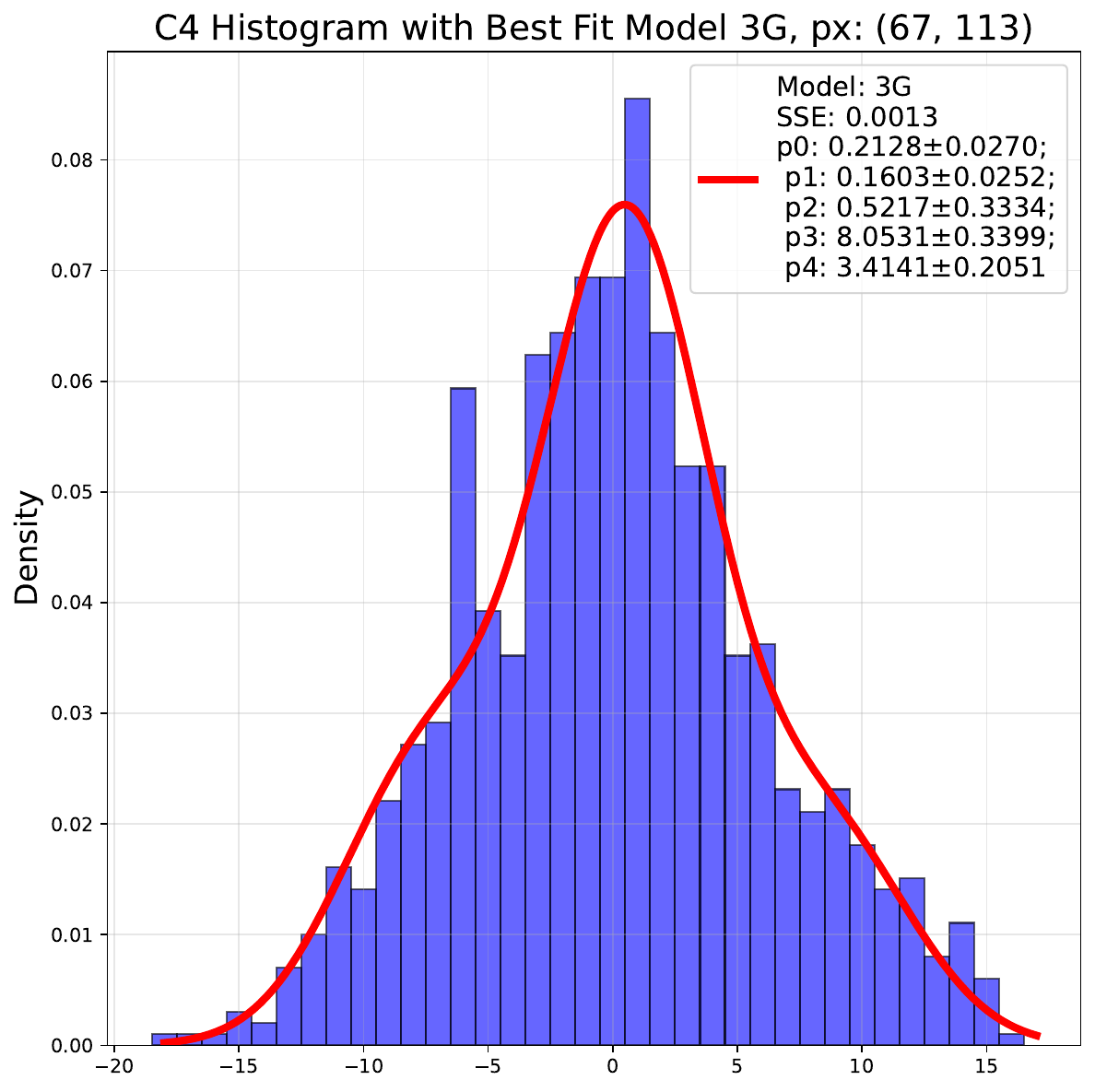} \\
  \end{tabular}
  \caption{\label{fig:c4_fits} Random examples of fits for C4 sensor.}
\end{figure}

The prepared mixture of Gaussians functions contain the following properties to reduce the amount of parameters to be fitted and to improve the fit quality.
\begin{itemize}
    \item Each function contains a center Gaussian, where most of the pixel output is expected to be, therefore containing the most weight. The side Gaussians are called sidelobes.
    \item The sidelobe means for 3G, and 5G, are expected to be symmetrical around the center mean and, therefore calculated using the mean shift parameter. For 5G, two different mean shifts are supplied. 
    
    \item For a given mixture level (3G, 5G), the Gaussian components contained in that level are expected to have the same or similar standard deviations due to RON, therefore the standard deviations are tied for all the Gaussian components in that level.
    \item For higher mixture levels (3G and 5G), the center weight component is calculated from the input sidelobe weights.
\end{itemize}

\textit{The amount of Gaussians in the mixture, can be scaled as much as needed. However, one should be careful because in our experiments, in most cases a higher level of mixture results in a better fit. Therefore, preliminary visual inspection is preferred to adjust the fit according to your data for the best results.}

After all the fits are done, the sum of squared errors (SSE) is calculated for each fit. The fit with the best SSE is selected as the best model.

\begin{figure}[h]
  \centering
  \begin{tabular}{ccc}
    \includegraphics[width=0.25\textwidth]{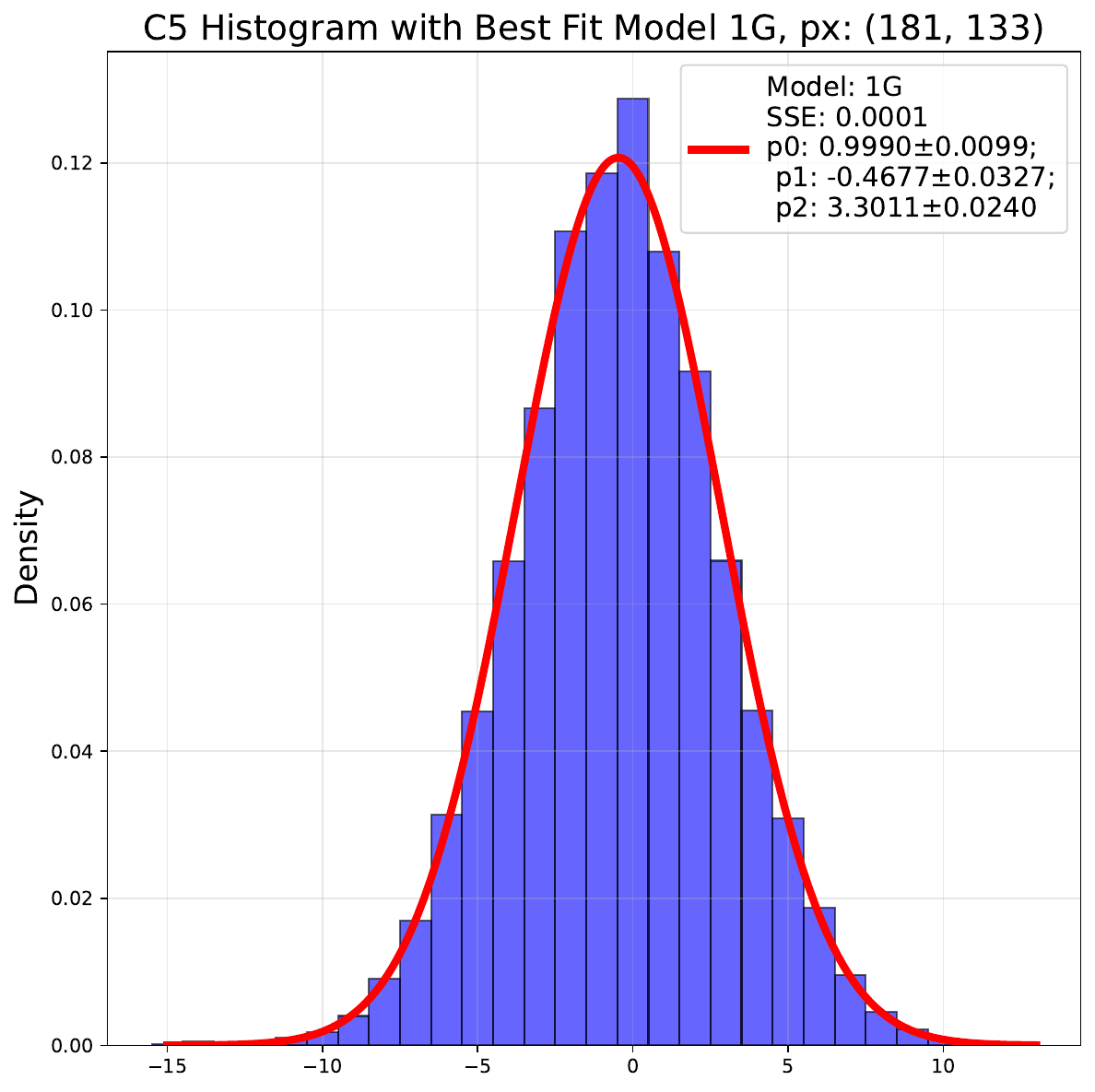} & 
    \includegraphics[width=0.25\textwidth]{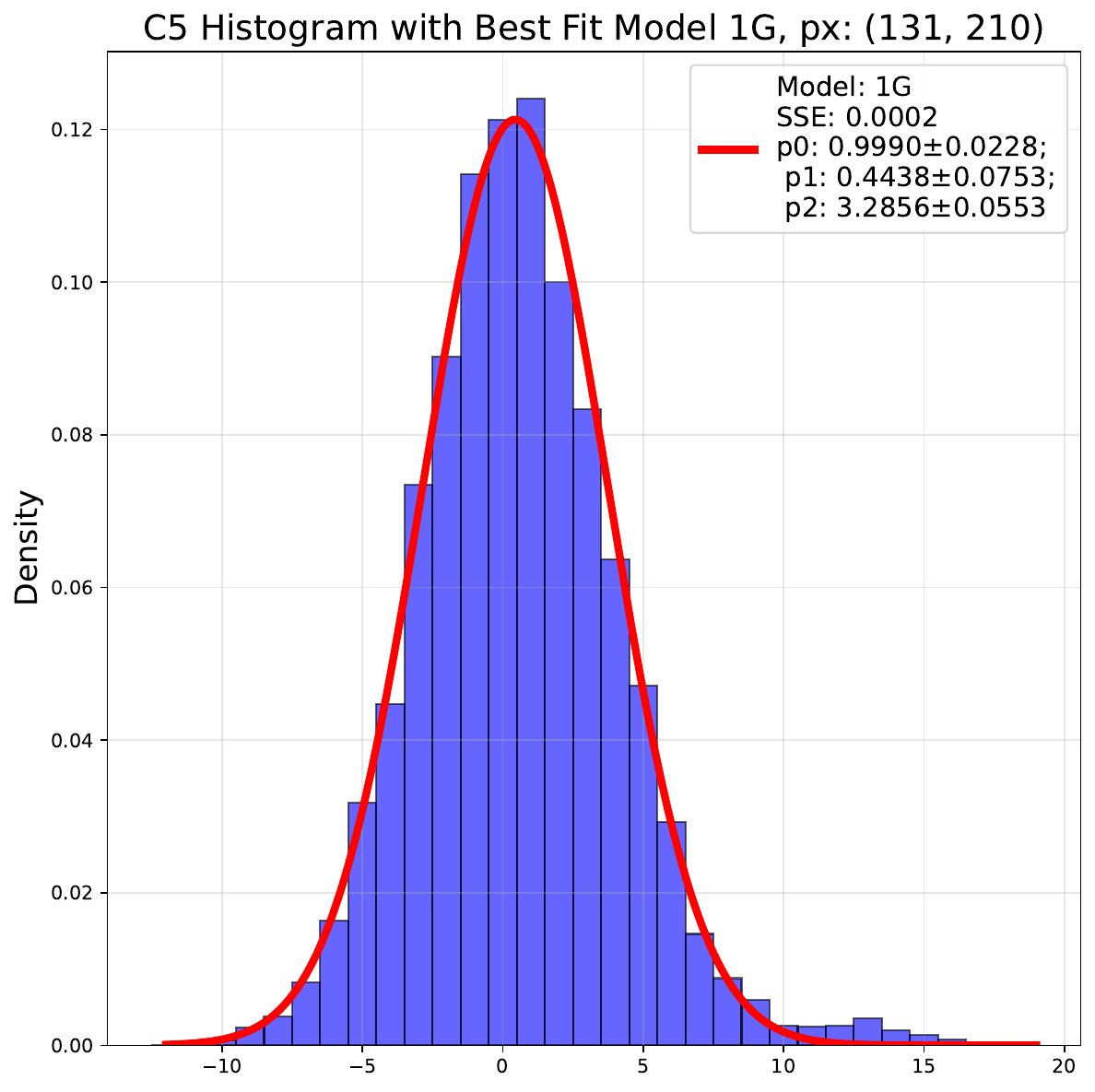} & 
    \includegraphics[width=0.25\textwidth]{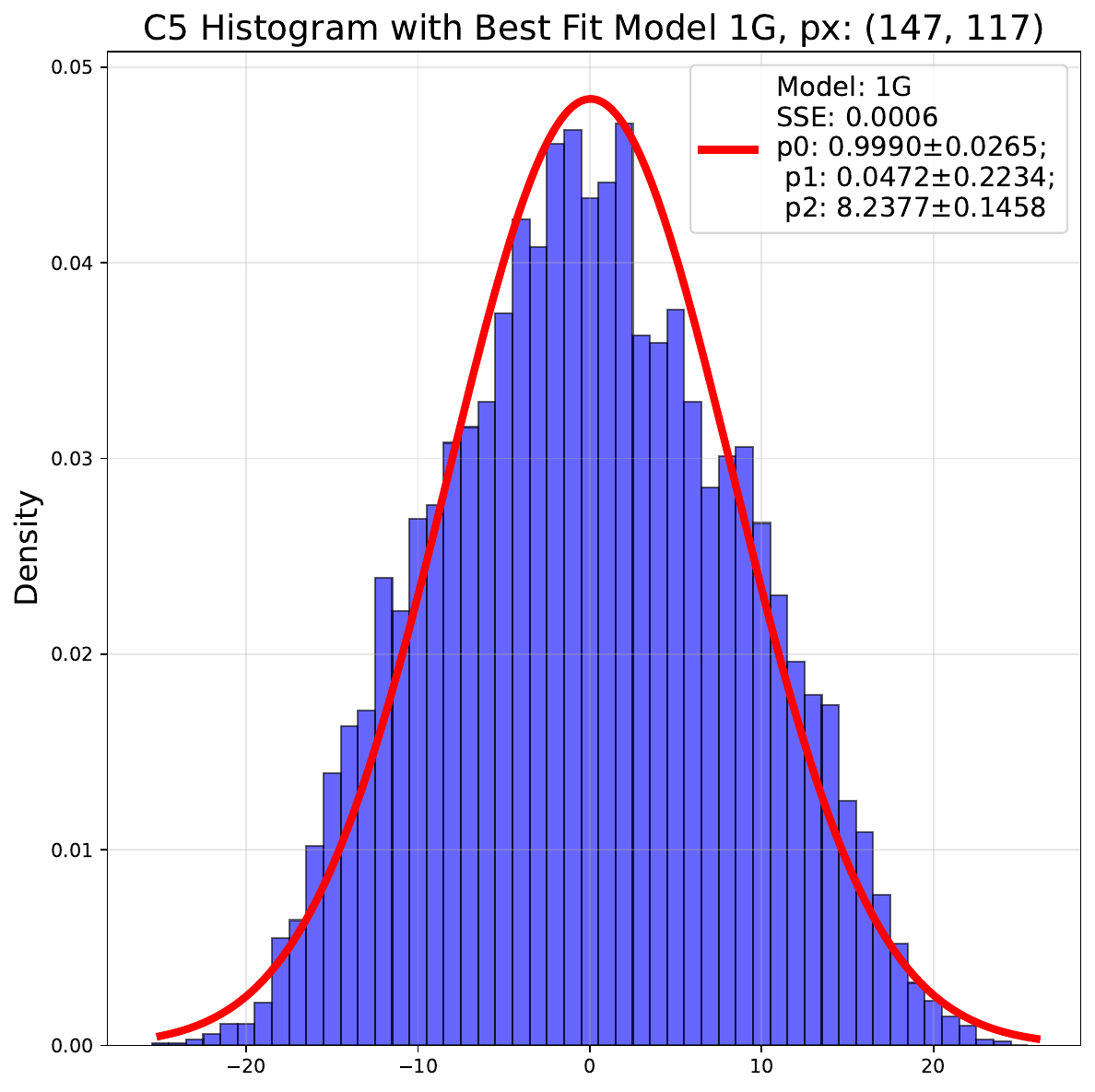} \\
    \includegraphics[width=0.25\textwidth]{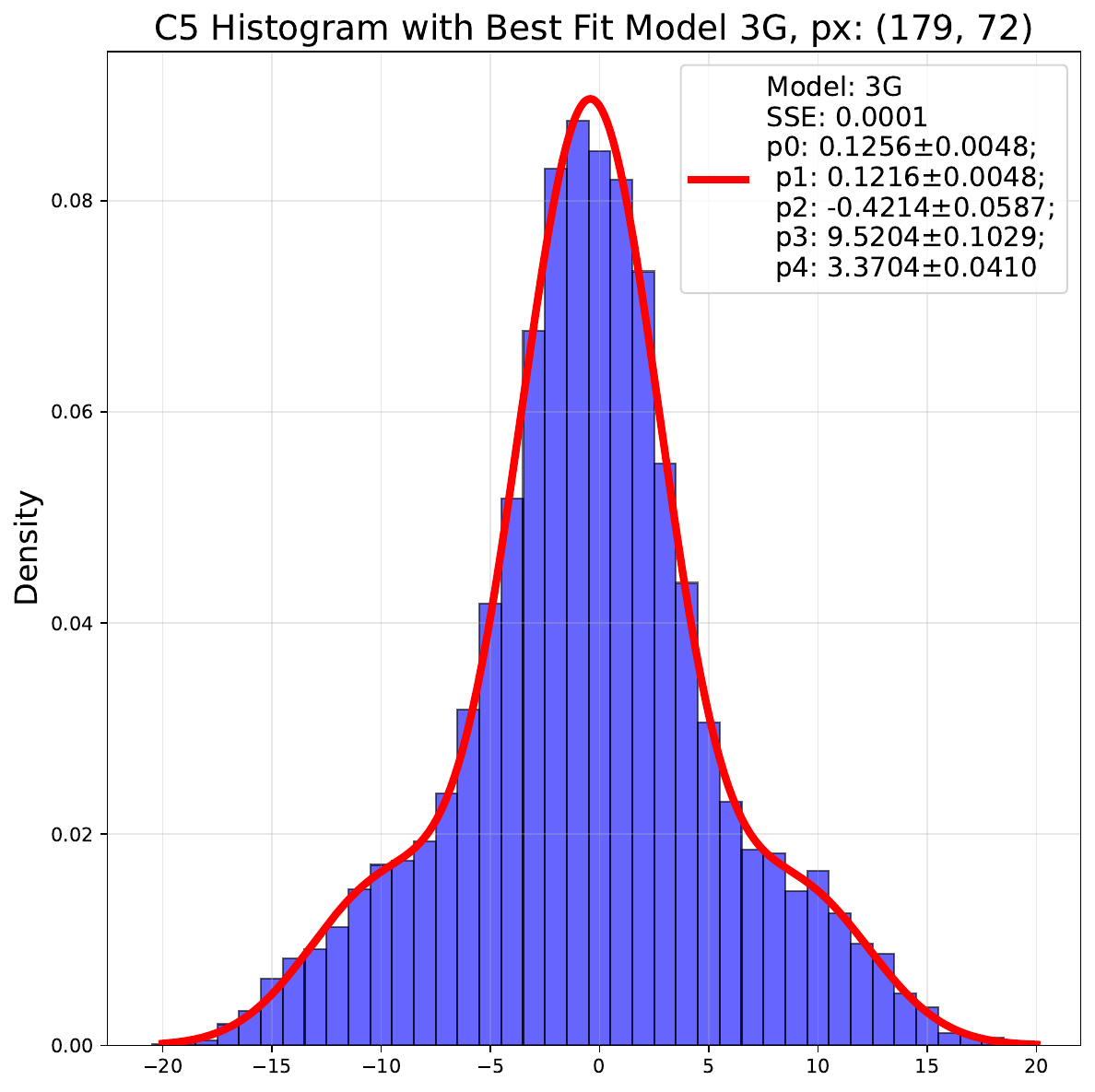} & 
    \includegraphics[width=0.25\textwidth]{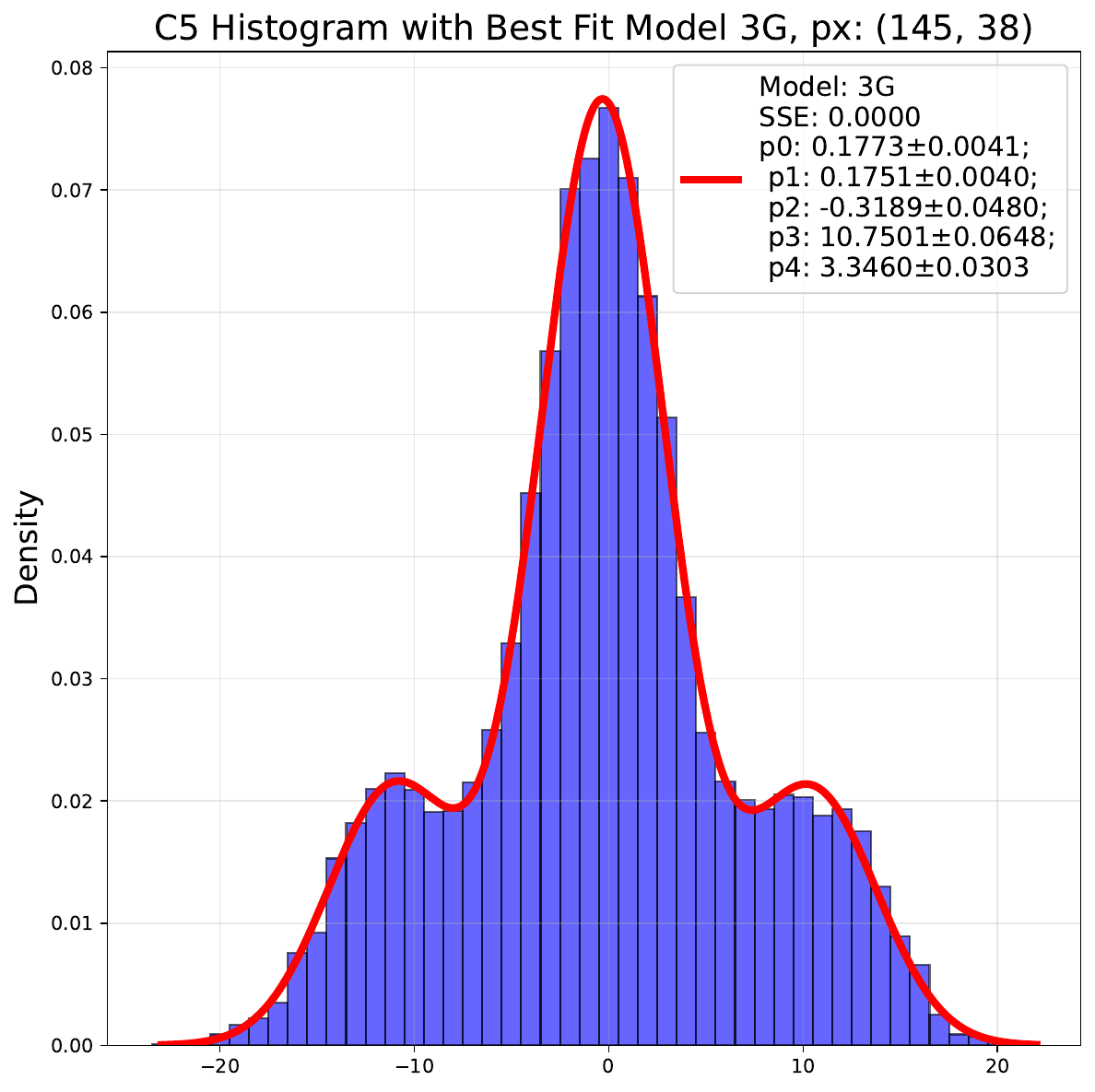} & 
    \includegraphics[width=0.25\textwidth]{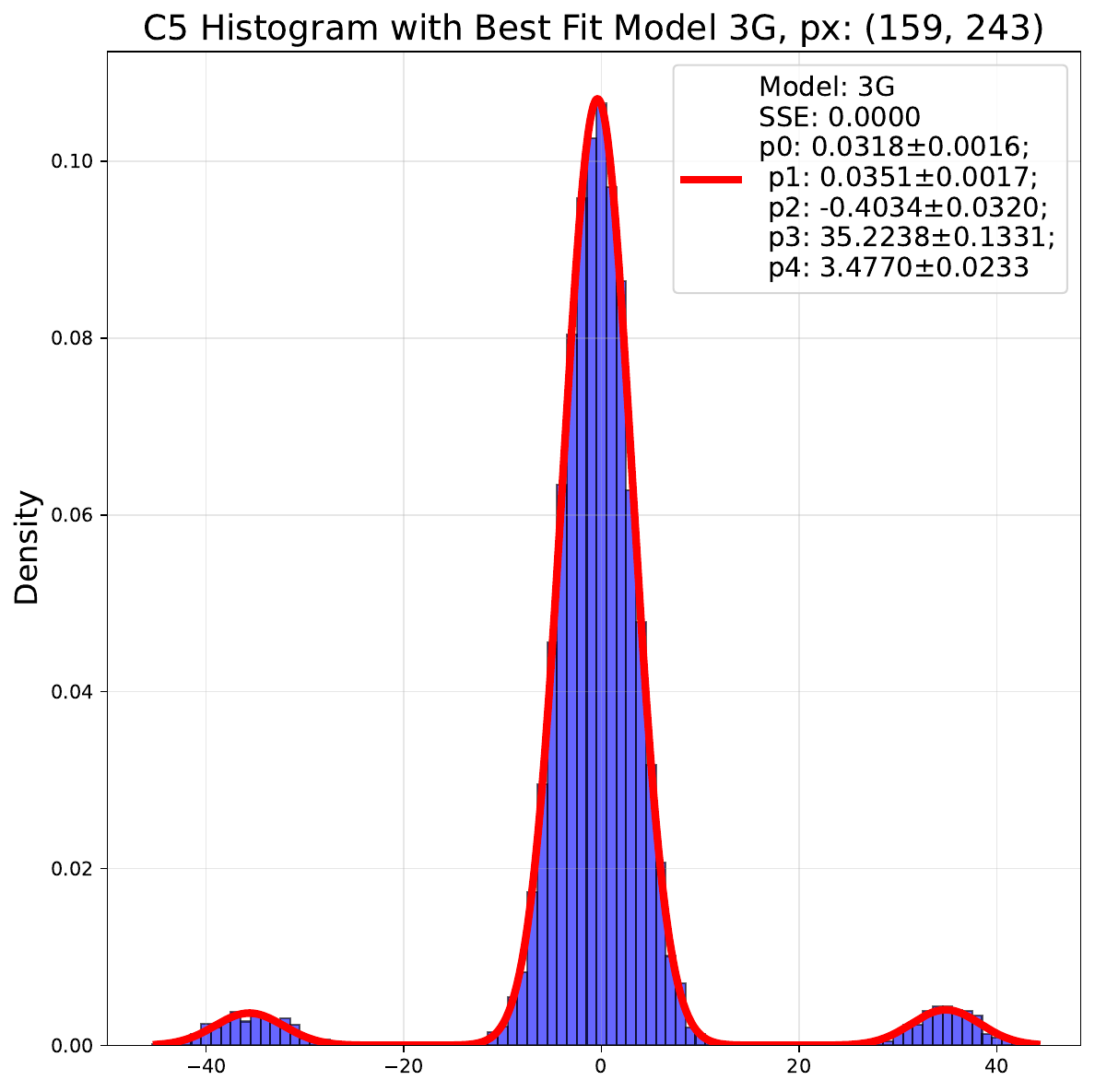} \\
    \includegraphics[width=0.25\textwidth]{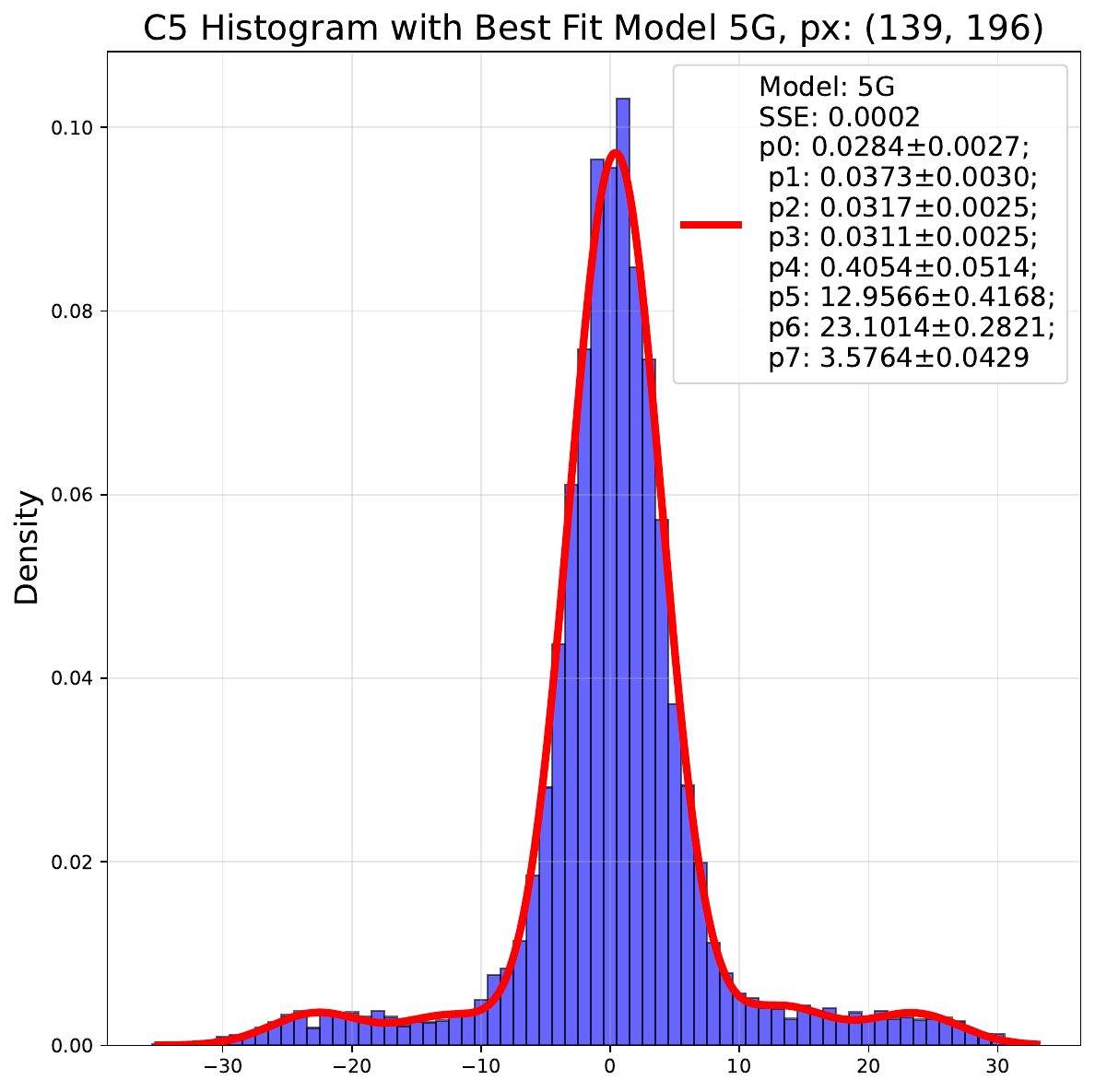} & 
    \includegraphics[width=0.25\textwidth]{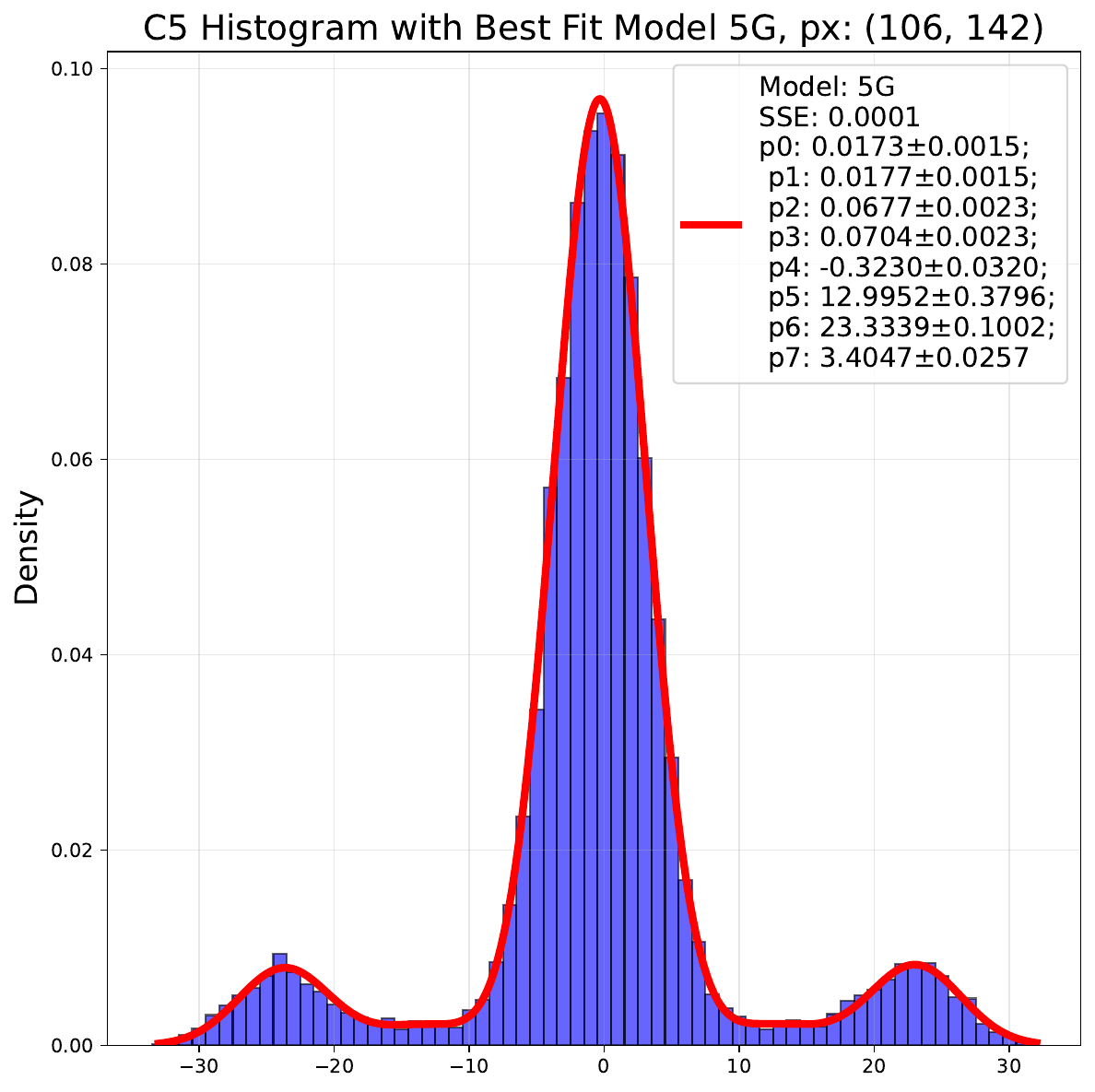} & 
    \includegraphics[width=0.25\textwidth]{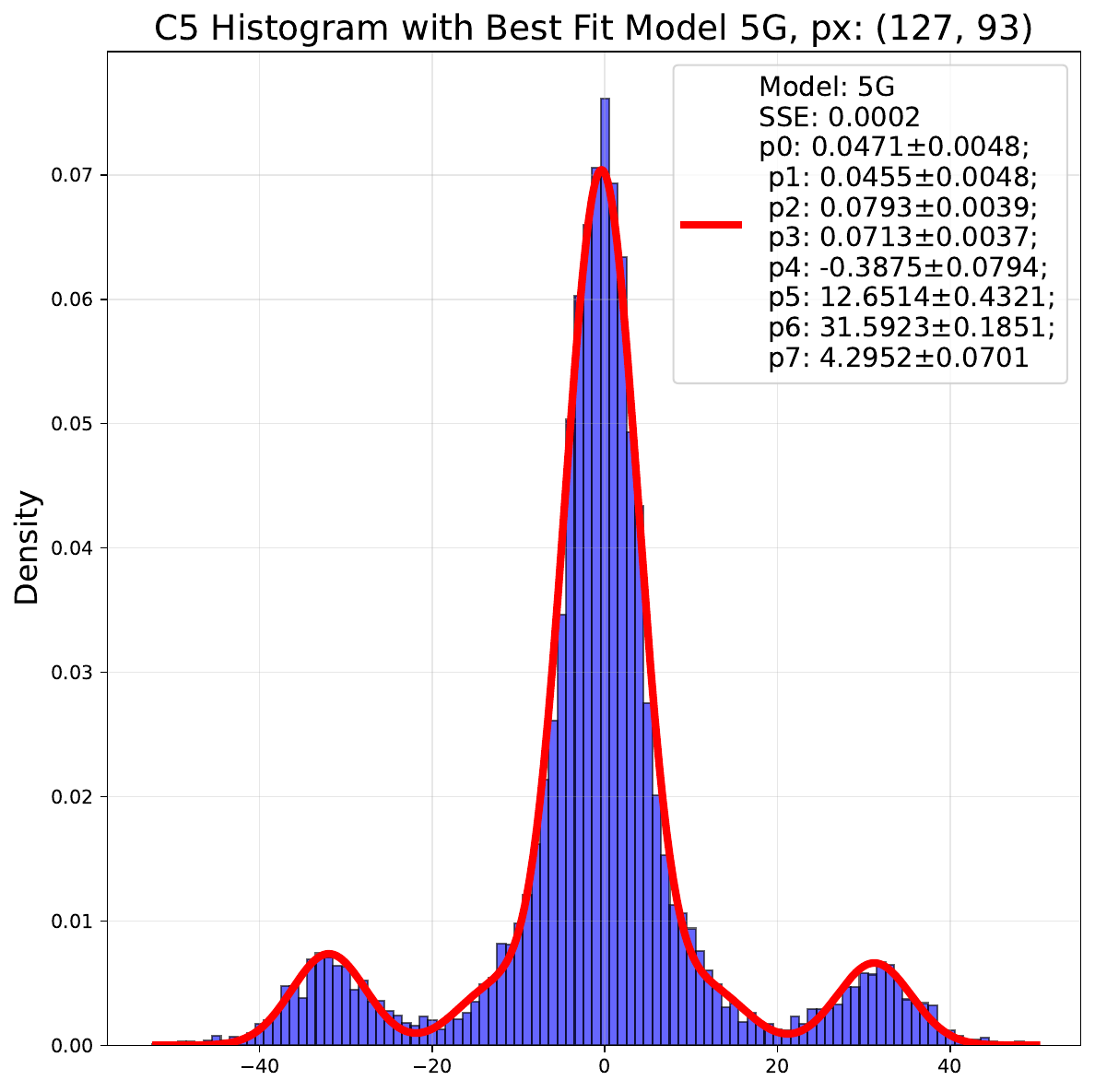} \\
  \end{tabular}
  \caption{\label{fig:c5_fits} Random examples of fits for C5 sensor.}
\end{figure}

\subsection{Correction}
During our initial trials, it was seen that even in the pixels which doesn't show any observable RTN, the best fit is selected as higher level of a mixture of Gaussians instead of a single normal distribution, partly due to SSE being calculated does not incorporate parameter estimation errors, hence, resulting in a lower value for higher level of mixtures. Moreover, small deviations from a perfect normal distribution can result in higher mixtures simply fitting better. In order to increase the accuracy of the fit, this has to be corrected.

The correction method depends on an iterative threshold based rejection and subsequent re-fitting. The threshold is calculated in two different ways as follows:
\begin{itemize}
    \item \textbf{Error Outlier Threshold: } 
    The standard deviation errors for each parameter for a given model for all pixels' best models are accumulated. The median and the standard deviation are calculated separately for each of the parameters. The best model of the pixels for which any parameter error exceeds the range of $3\sigma$ is rejected. 
    \item \textbf{Error/Parameter Ratio: } For each pixel's best model (best fit), the standard deviation error to parameter value ratio is calculated. If this ratio exceeds 20\% the fitted model is rejected. 
\end{itemize}
and the pixels that are rejected by either of the thresholds are refitted using a lower level of mixture (5G $\rightarrow$ 3G $\rightarrow$ 1G). The pixels with 1G fits are not further checked. This process is iteratively followed until less than 10 pixels are refitted.

\section{Results and Discussion}\label{sec:res}

\begin{figure}[ht]
  \centering
  \begin{tabular}{c}
    \includegraphics[width=\textwidth]{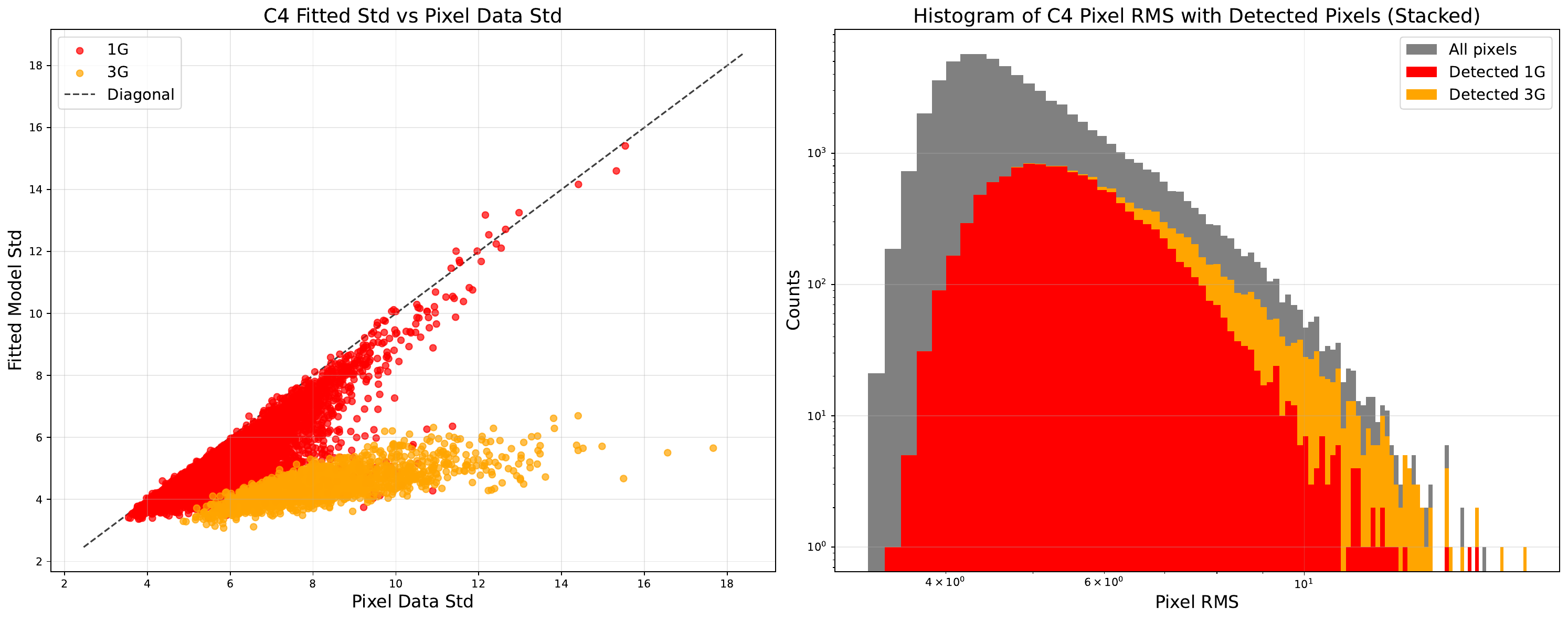} 
  \end{tabular}
  \caption{\label{fig:c4_hist_std} On the left, the fitted Gaussians' standard deviation against pixel standard deviation and on the right, stacked histogram of fitted models' standard deviations overlayed on top of all pixels' RMS for C4.}
\end{figure}

\begin{figure}[ht]
  \centering
  \begin{tabular}{c}
    \includegraphics[width=\textwidth]{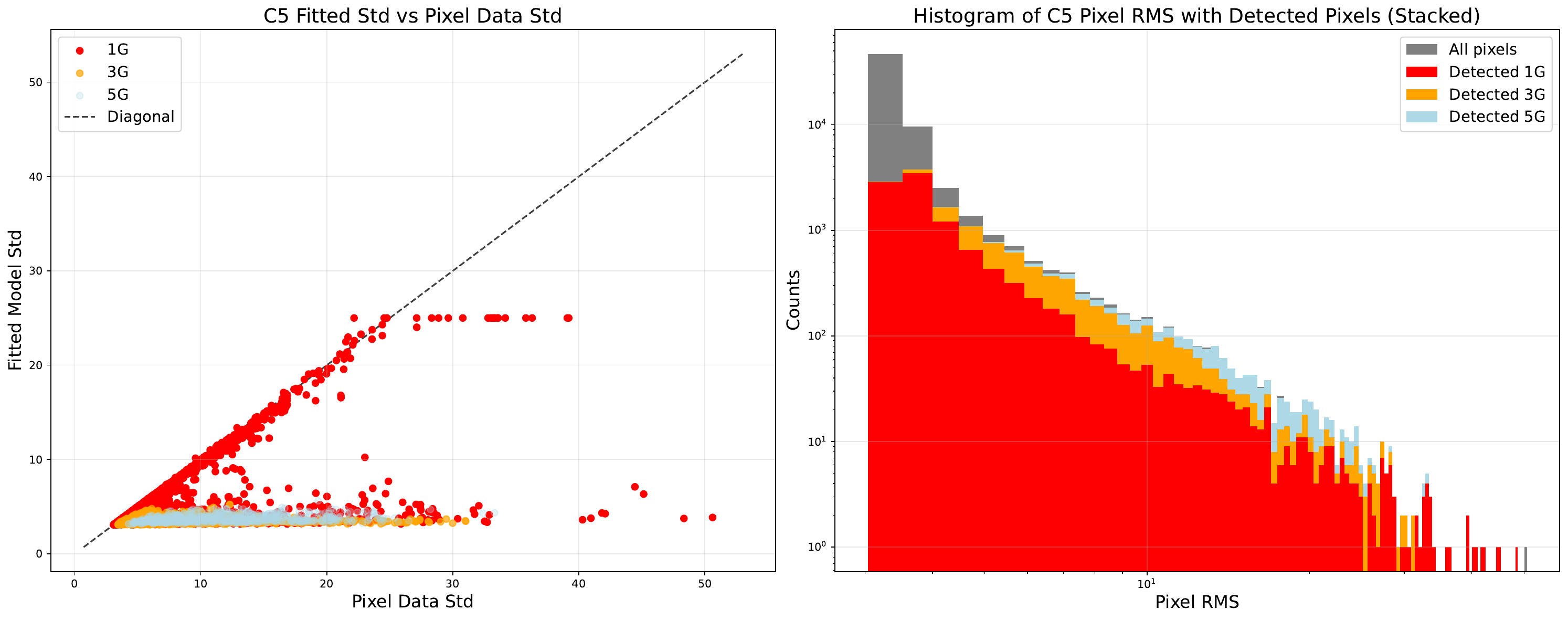} 
  \end{tabular}
  \caption{\label{fig:c5_hist_std} On the left, the fitted Gaussians' standard deviation against pixel standard deviation, and on the right, stacked histogram of fitted models' standard deviations overlayed on top of all pixels' RMS for C5.}
\end{figure}

The statistics for both cameras within the tested region is given below: 
\begin{itemize}
    \item[]  \textbf{For C4:}
    \item $256\times256 = 65536$ pixels are analyzed.
    \item Within those pixels, $14432 \simeq 22.0\%$ of them are completely rejected by the AD normality test.
    \item Out of the rejected pixels, $1927 \simeq 2.94\%$  characterized as 3G, and rest are fitted with 1G.
\end{itemize}

\begin{itemize}
    \item[]  \textbf{For C5:}
    \item $256\times256 = 65536$ pixels are analyzed.
    \item Within those pixels, $14530 \simeq 22.2\%$ of them are completely rejected by the AD normality test.
    \item Out of the rejected pixels, $706 \simeq 1.08\%$ characterized as 5G, $3340 \simeq 5.10\%$ characterized as 3G, and rest are fitted with 1G.
\end{itemize}
Some examples of fits of each model for C4 and C5 are shown in Fig.~\ref{fig:c4_fits} and Fig.~\ref{fig:c5_fits}, respectively. The resulting percentage of fits is in alignment with the expected range which is below $5\%$ for both sensors. Both cameras show a range of variable RTN characteristics, resulting in different levels and weights of jumps. The algorithm performs well under a broad range of RTN peaks. However, limitations of the fitting and evaluation algorithm show some possible errors.

In Fig.~\ref{fig:c4_hist_std} and in Fig.~\ref{fig:c5_hist_std}, the standard deviation and the stacked histogram of C4 and C5 are shown, respectively. When the fitted versus actual standard deviations of the pixels are analyzed, in a perfectly working algorithm, all the 1G fitted pixels should have lied on the diagonal. However, the spread of those diagonals shows that the algorithm fails to detect some RTN cases. The hard limitation of standard deviation for the C5 sensor is also observable, however, this is not a big concern.
From the stacked histograms, it is clear that the RTN presents itself throughout the whole RMS range, hence blocking a sensor-wide statistics-based detection. 

\begin{figure}[b]
  \centering
  \begin{tabular}{c}
    \includegraphics[width=0.7\textwidth]{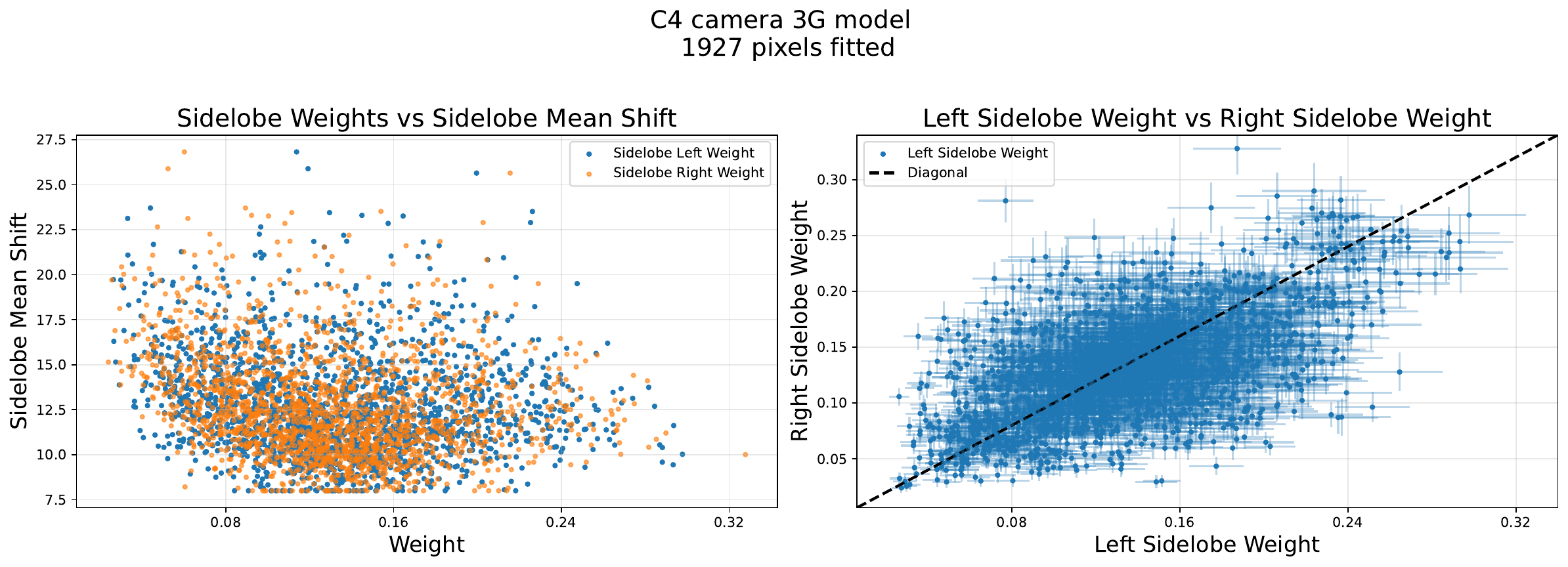} 
  \end{tabular}
  \caption{\label{fig:c4_3g_weight_plots} Resulting 3G fits data for C4. Left: Sidelobe weights against sidelobe mean shifts. Right: Weight comparison for left and right sidelobes.}
\end{figure}

\begin{figure}[b]
  \centering
  \begin{tabular}{c}
    \includegraphics[width=0.7\textwidth]{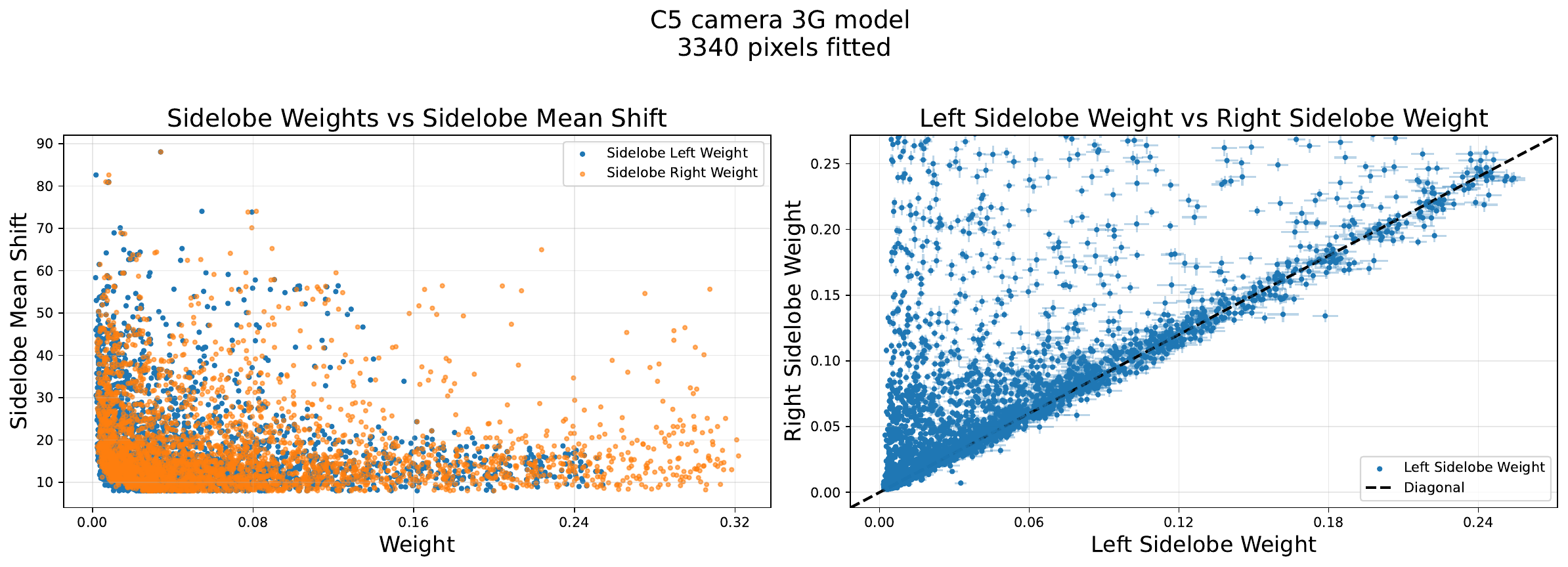} 
  \end{tabular}
  \caption{\label{fig:c5_3g_weight_plots} Resulting 3G fits data for C5. Left: Sidelobe weights against sidelobe mean shifts. Right: Weight comparison for sidelobes.}
\end{figure}

\begin{figure}[t]
  \centering
  \begin{tabular}{c}
    \includegraphics[width=0.7\textwidth]{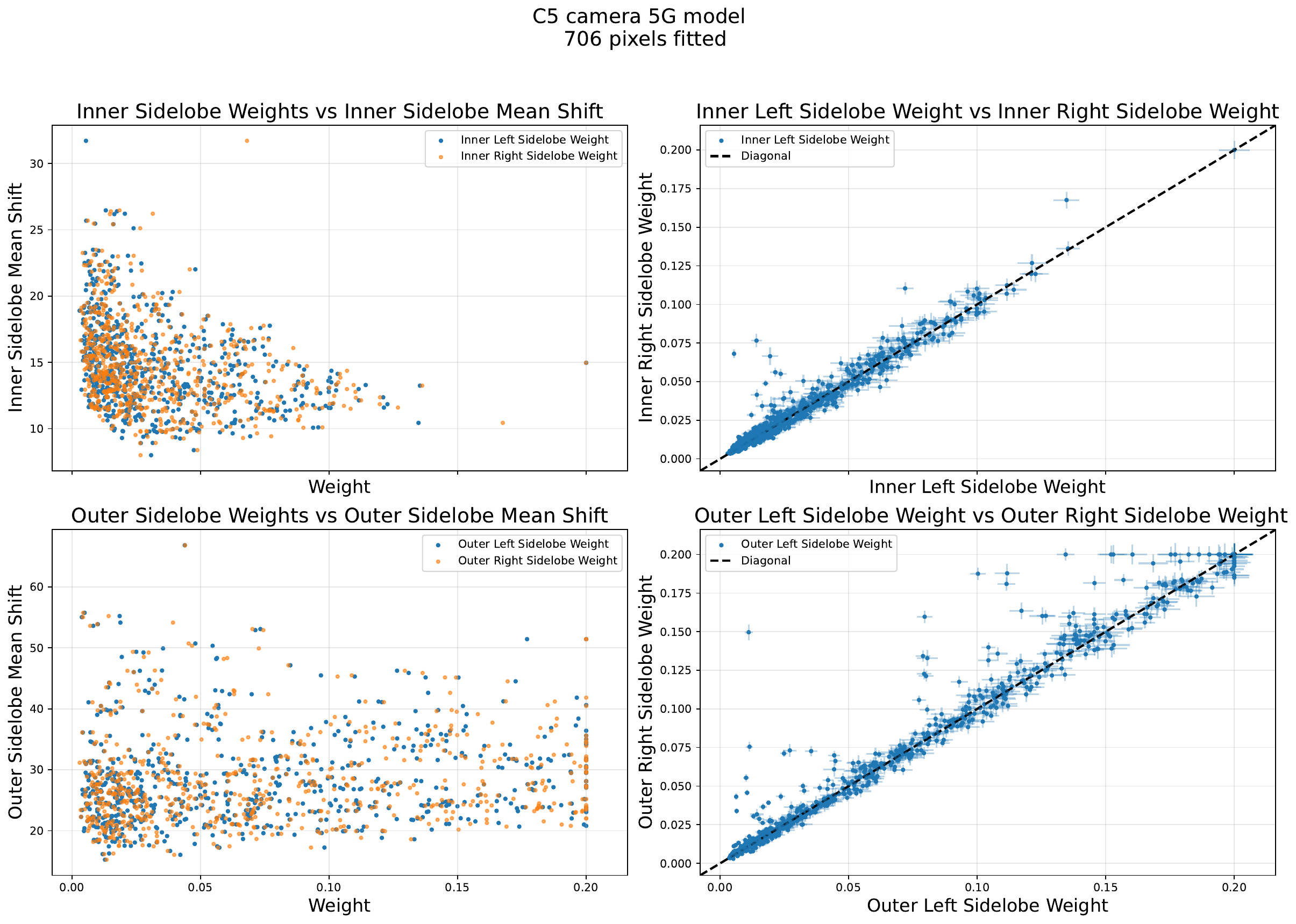} 
  \end{tabular}
  \caption{\label{fig:c5_5g_weight_plots} Resulting 5G fits data for C5. Top left: Inner sidelobe weights against inner sidelobe mean shifts. Top right: Weight comparison for inner sidelobes. Bottom left: Outer sidelobe weights against outer sidelobe mean shifts. Bottom right: Weight comparison for outer sidelobes. }
\end{figure}

In Fig.~\ref{fig:c4_3g_weight_plots}, Fig.~\ref{fig:c5_3g_weight_plots}, and Fig.~\ref{fig:c5_5g_weight_plots}, the fitted sidelobe weights and sidelobe mean shifts and sidelobe within themselves are correlated. While C4 clearly shows clustering around certain mean shifts and weights for 3G,  no definitive observation can be made for C5 3G fits. This can result from a relatively homogenous distribution of trap depths for C4. Still, C5 shows a similar clustering for 5G fits, especially at the lower-weight region. The hard limitation of the upper weight to eliminate overshooting of the fit around the corners can also be observed. 
However, for both sensors, the instability of the fitting function below the weights of $10^{-2}$ is prominent, as there are no fitted data. Hence making the algorithm miss or over-estimate some of the possible RTN pixels. 
When the symmetrical sidelobe ratios are analyzed, it is observed that the C4 follows our initial assumption of symmetrical jumps, however, incorporates a large variation. For C5 3G, one can see that there is a higher probability of a positive jump compared to a negative jump due to RTN. This means that more electrons are released or tunneled into the accumulation rather than being captured by the traps, which can be due to the smaller pixel size of the sensor.

\begin{figure}[hb]
  \centering
  \begin{tabular}{c}
    \includegraphics[width=\textwidth]{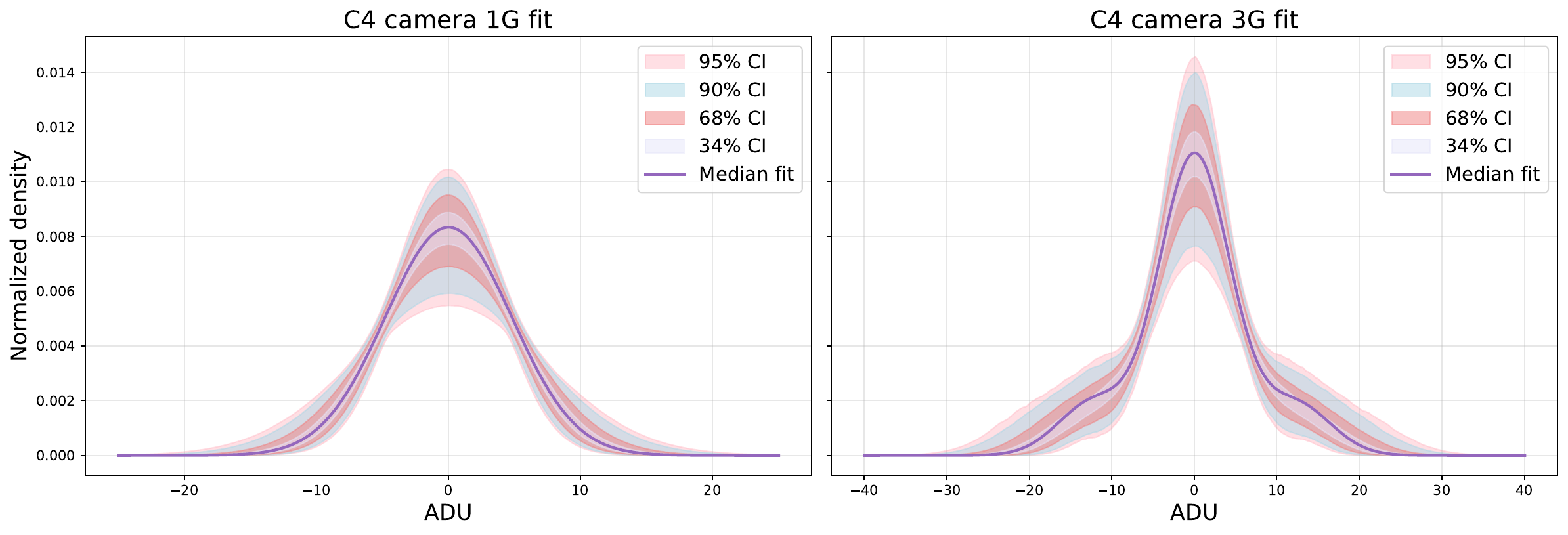} 
  \end{tabular}
  \caption{\label{fig:c4_empiric} The distribution of fits based on actual fitted data for C4. CI represents the confidence intervals and the median fit is the fit that corresponds to the median selection of each parameter for the given model.}
\end{figure}

\begin{figure}[ht]
  \centering
  \begin{tabular}{c}
    \includegraphics[width=\textwidth]{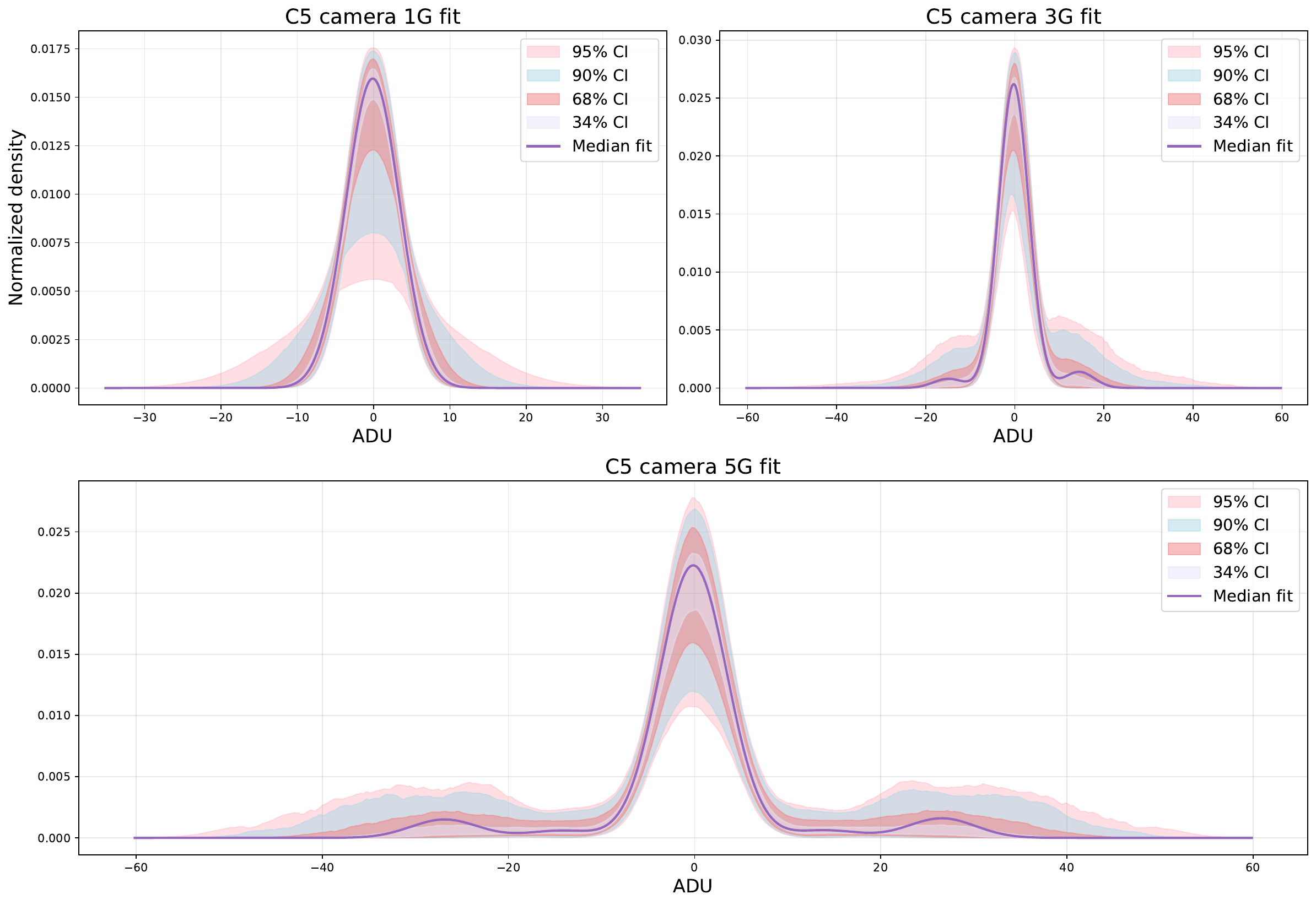} 
  \end{tabular}
  \caption{\label{fig:c5_empiric} The distribution of fits based on actual fitted data for C5. CI represents the confidence intervals and the median fit is the fit that corresponds to the median selection of each parameter for the given model.}
\end{figure}

In Fig.~\ref{fig:c4_empiric} and in Fig.~\ref{fig:c5_empiric}, empirical fit confidence intervals are constructed, from all the fits for each model. These plots give an idea of what an approximate pixel output histogram can look like. The unequal weights for left and right sidelobes are visible in Fig.~\ref{fig:c5_empiric} for C5 3G. For higher-level fits, it is observable that the most prominent initial peak after the center peak is around 10-20 ADU and the second one is between 20-40 ADU.

\section{Conclusion}\label{sec:conc}

In this work, we presented a systematic method for detecting and characterizing random telegraph noise (RTN) in sCMOS image sensors only using a series of dark frames. A dithering approach for preparing the integer rounded data for normality tests is applied. To the best of our knowledge, no such method was found in the literature. By combining the Anderson-Darling statistical normality test and Gaussian‐mixture fitting on pixel histogram, we demonstrated a practical algorithm that flags RTN‐affected pixels and subsequently models their multiple discrete levels. Empirical evaluations of two sCMOS cameras, namely C4 and C5, showed that only a small fraction (1–5\%) of pixels display RTN which is in alignment with the previous literature. Our results indicate that RTN pixels exhibit a variety of characteristics suggesting the need for careful parameter initialization and fitting constraints. For C4, the first peak's mean shift is mostly contained within $\pm 10$ to $15$ ADUs of the median of the pixel's output, and for C5, the first peak is around $\pm 15$ ADUs, and the second peak is concentrated within $\pm 20$  to $\pm 30$ ADUs around the median of the pixel's output. Additionally, the derived confidence intervals for various mixture models highlight the range of possible RTN amplitudes and weights in C4 and C5 sensors. Future work includes refining the mixture‐fitting procedure to mitigate overfitting in high‐mixture models and investigating on‐the‐fly or real‐time calibration approaches that automatically correct RTN contamination in sCMOS detectors during acquisition. Ultimately, improved RTN detection and correction will help ensure that sCMOS cameras continue to deliver high‐precision imaging across a range of scientific applications.

\acknowledgments 
 
This work 
is partly funded by the EU and supported by the Czech Ministry of Education, Youth and Sports through the project CZ.02.01.01/00/22\_008/0004596 (SENDISO). It is also partly funded by the Grant Agency of the Czech Technical University in Prague, grant No. SGS23/186/OHK3/3T/13.

\bibliography{report} 
\bibliographystyle{spiebib} 

\end{document}